%% file: arenamain.tex
\documentclass[sigconf]{acmart}
\renewcommand\footnotetextcopyrightpermission[1]{}
\usepackage{booktabs}
\usepackage{verbatim}
\usepackage{graphicx}
\usepackage{multirow}
\usepackage{array}
\usepackage{bm}
\usepackage{geometry}
\usepackage{amsmath}
\usepackage{amsfonts}
\usepackage{scalerel}
\usepackage[ruled]{algorithm2e}
\usepackage{algorithmic}
\usepackage{caption}
\usepackage{subcaption}
\usepackage{multicol}
\usepackage{color}
\usepackage{epsfig}
\usepackage{natbib}
\usepackage{xspace}
\usepackage{booktabs}
\usepackage{url}
\usepackage{enumitem}
\usepackage{balance}
\graphicspath{{figure/}}

\newcommand{\hide}[1]{} 
\newcommand{\vpara}[1]{\vspace{0.05in}\noindent\textbf{#1 }}
\newcommand{\para}[1]{\vspace{0.05in}\noindent\textbf{#1 }}
\newcommand{\secref}[1]{\S\ref{#1}}

\newcommand{\figref}[1]{Fig.~\ref{#1}}
\newcommand{\tableref}[1]{Table~\ref{#1}}

\newcommand{\kingofglory}{{\sf Honor of Kings}\xspace}

\newcommand{\warrior}{\,\vcenter{\hbox{\ensuremath{%
  \mathchoice{\includegraphics[height=3ex]{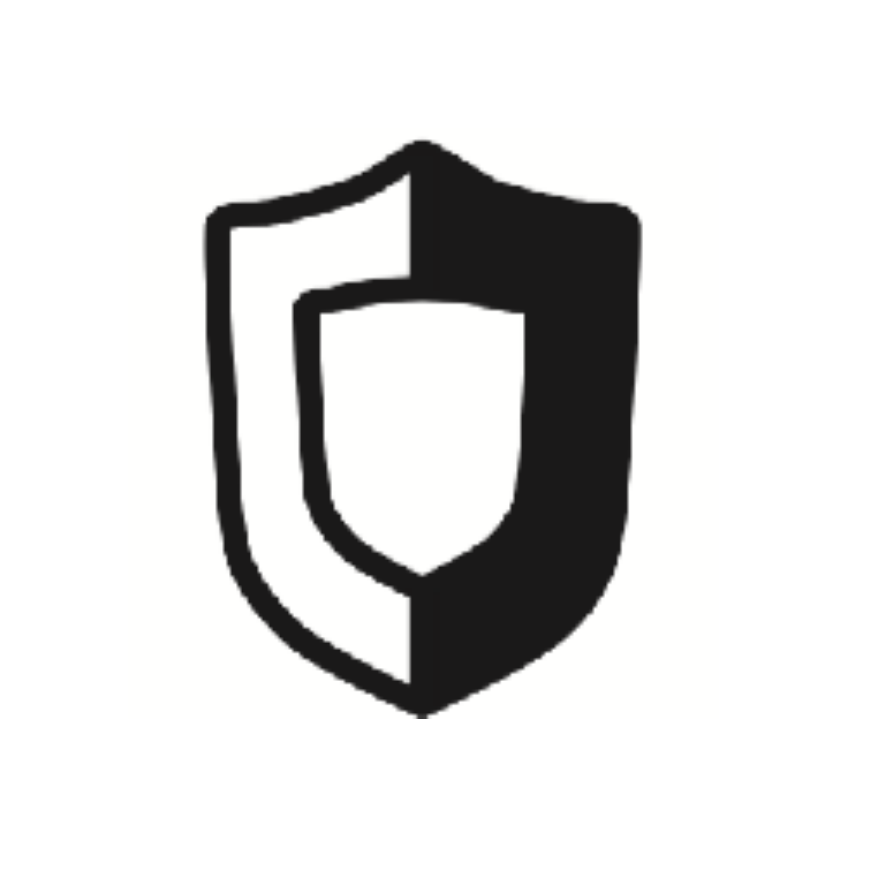}}
    {\includegraphics[height=3ex]{figure/warrior.pdf}}
    {\includegraphics[height=2.5ex]{figure/warrior.pdf}}
    {\includegraphics[height=2ex]{figure/warrior.pdf}}
}}}\,}

\newcommand{\mage}{\,\vcenter{\hbox{\ensuremath{%
  \mathchoice{\includegraphics[height=3ex]{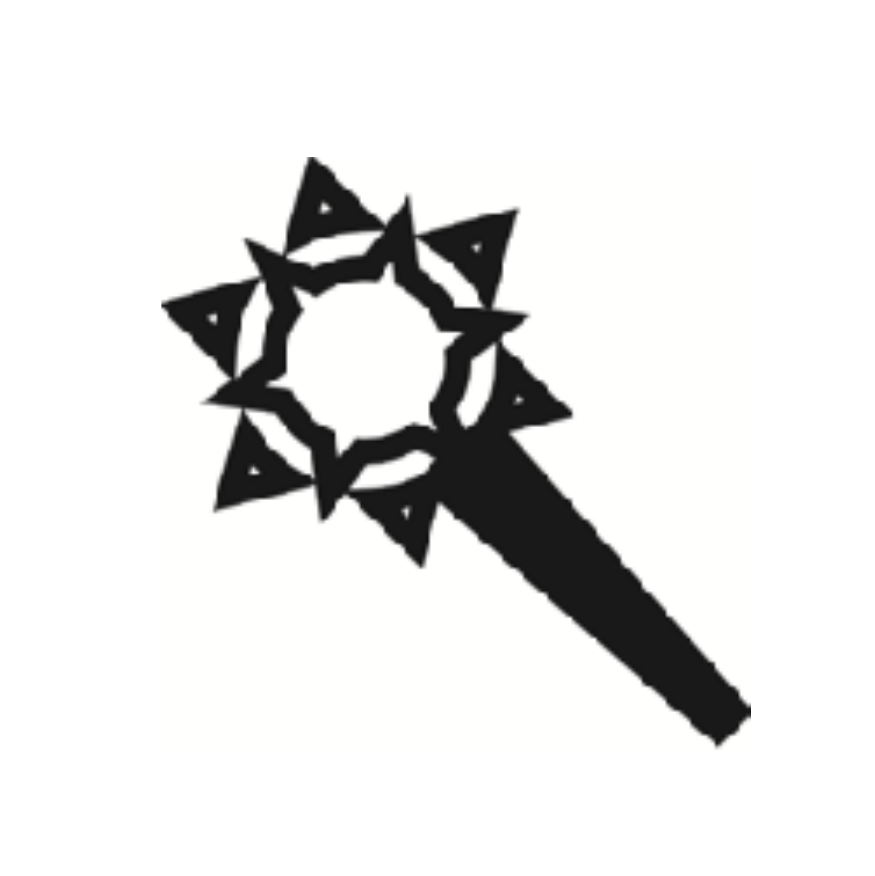}}
    {\includegraphics[height=3ex]{figure/mage.pdf}}
    {\includegraphics[height=2.5ex]{figure/mage.pdf}}
    {\includegraphics[height=2ex]{figure/mage.pdf}}
}}}\,}

\newcommand{\marksman}{\,\vcenter{\hbox{\ensuremath{%
  \mathchoice{\includegraphics[height=3ex]{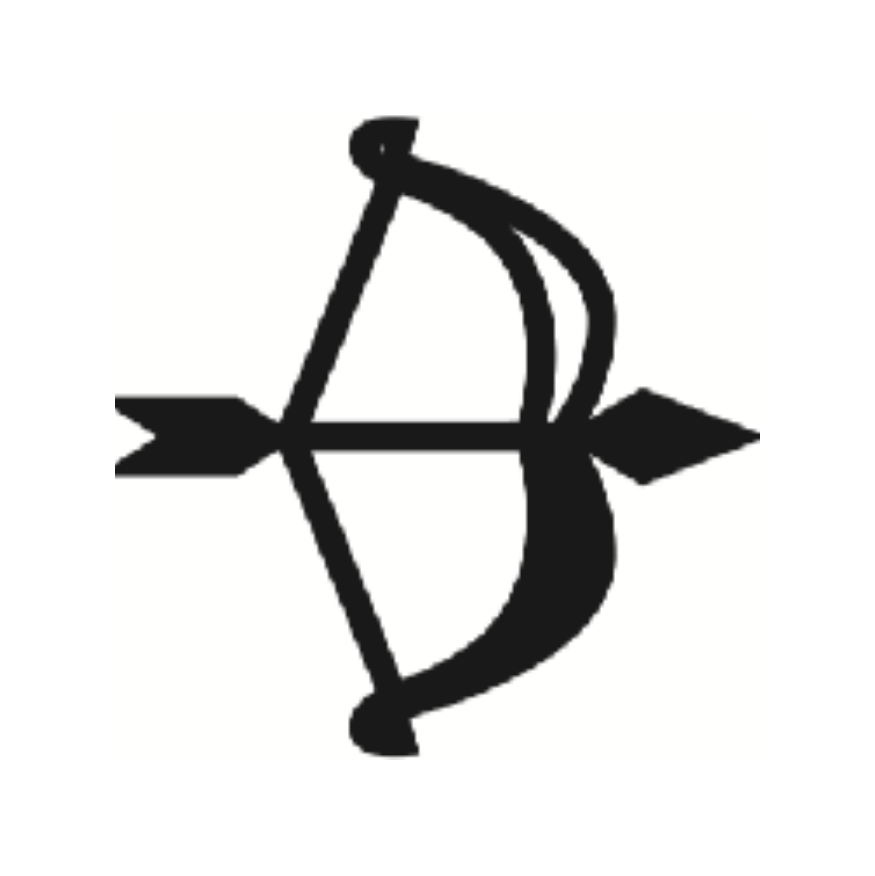}}
    {\includegraphics[height=3ex]{figure/marksman.pdf}}
    {\includegraphics[height=2.5ex]{figure/marksman.pdf}}
    {\includegraphics[height=2ex]{figure/marksman.pdf}}
}}}\,}

\newcommand{\assassin}{\,\vcenter{\hbox{\ensuremath{%
  \mathchoice{\includegraphics[height=3ex]{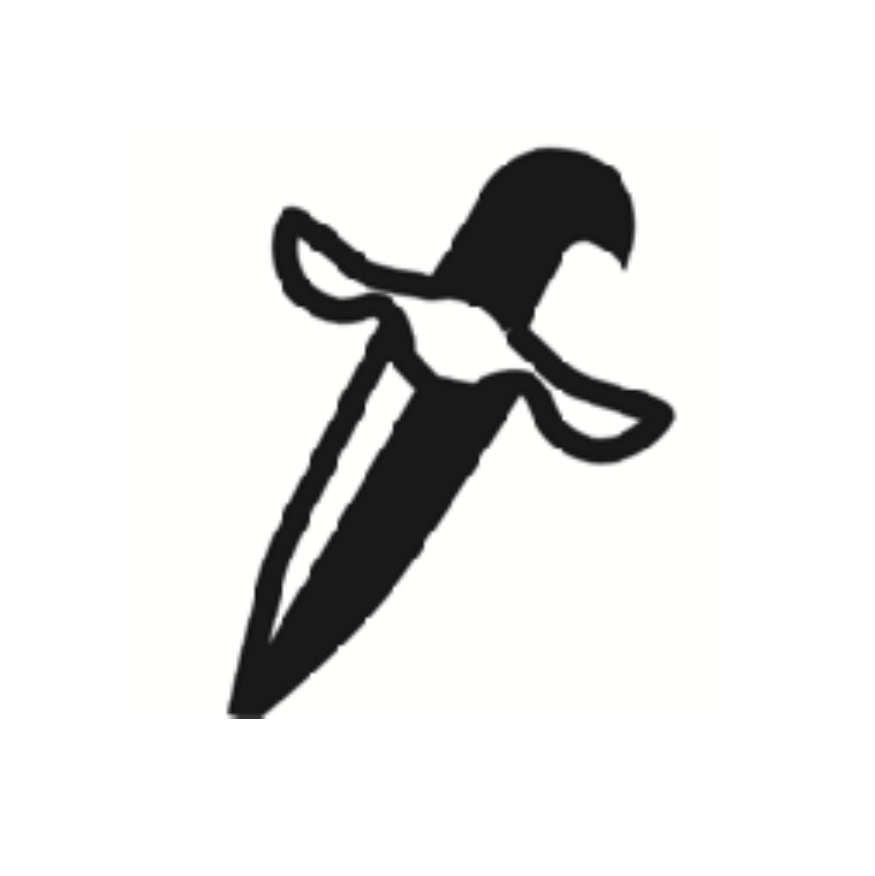}}
    {\includegraphics[height=3ex]{figure/assassin.pdf}}
    {\includegraphics[height=2.5ex]{figure/assassin.pdf}}
    {\includegraphics[height=2ex]{figure/assassin.pdf}}
}}}\,}

\newcommand{\support}{\,\vcenter{\hbox{\ensuremath{%
  \mathchoice{\includegraphics[height=3ex]{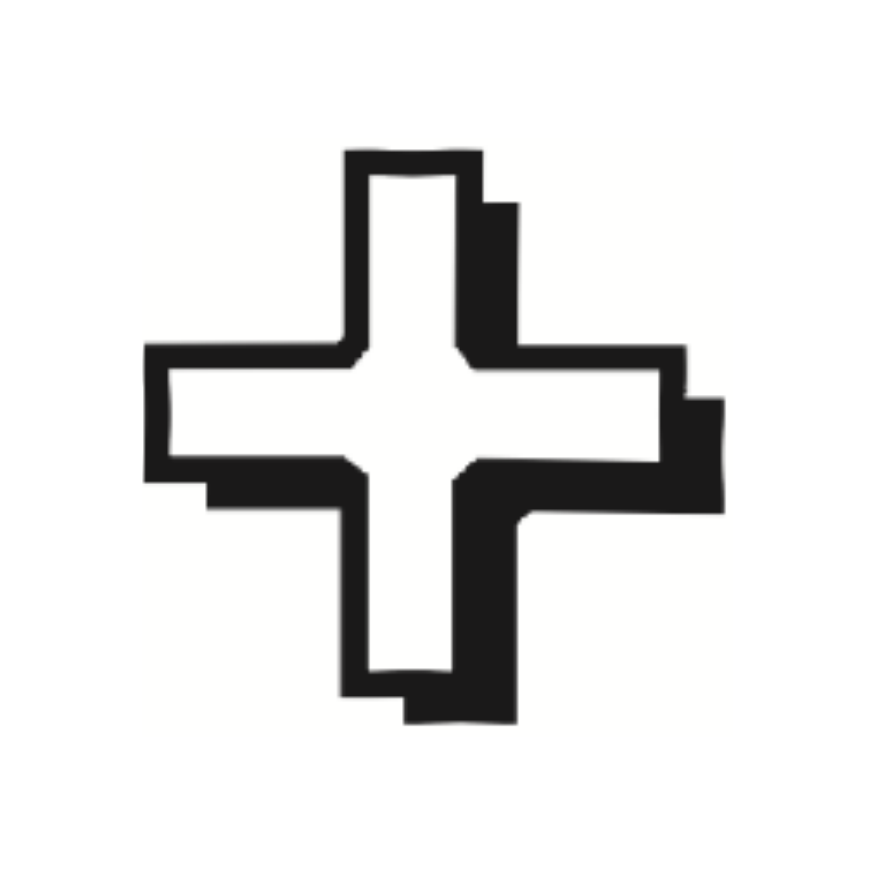}}
    {\includegraphics[height=3ex]{figure/support.pdf}}
    {\includegraphics[height=2.5ex]{figure/support.pdf}}
    {\includegraphics[height=2ex]{figure/support.pdf}}
}}}\,}

\begin{document}
\title{
What Makes a Good Team? 
A Large-scale Study \\
on the Effect of Team Composition in Honor of Kings}

\author{Ziqiang Cheng}
\affiliation{
  \institution{Zhejiang University}
}
\email{petecheng@zju.edu.cn}

\author{Yang Yang}
\affiliation{
  \institution{Zhejiang University}
}
\email{yangya@zju.edu.cn}

\author{Chenhao Tan}
\affiliation{
  \institution{University of Colorado Boulder}
}
\email{chenhao@chenhaot.com}

\author{Denny Cheng}
\affiliation{
  \institution{Tencent Corporation}
}
\email{dennycheng@tencent.com}

\author{Yueting Zhuang}
\affiliation{
  \institution{Zhejiang University}
}
\email{yzhuang@zju.edu.cn}

\author{Alex Cheng}
\affiliation{
  \institution{Tencent Corporation}
}
\email{alexcheng@tencent.com}

\renewcommand{\shortauthors}{Cheng et al.}
\renewcommand{\shorttitle}{A Large-scale Study on the Effect of Team Composition in Honor of Kings}

\input{abstract.tex}



\maketitle

\input{intro}
\input{data}

\input{performance}
\input{surrender}
\input{abuse}
\input{exp}
\input{relate}
\input{conclude}

\balance 
\bibliographystyle{ACM-Reference-Format}
\bibliography{bibliography}

\input{appendix}

\end{document}

%% file: abstract.tex

\begin{abstract}
Team composition is a central factor in determining the effectiveness of a team.
In this paper, we present a large-scale study on the effect of team composition on multiple measures of team effectiveness.
We use a dataset from the largest multiplayer online battle arena (MOBA) game, \kingofglory, with 96 million matches involving 100 million players.
We measure team effectiveness based on team performance (whether a team is going to win), team tenacity (whether a team is going to surrender), and team rapport (whether a team uses abusive language).
Our results confirm the importance of team diversity and show that diversity has varying effects on team effectiveness:
although diverse teams perform well and show tenacity in adversity, they are more likely to abuse when losing than less diverse teams.
Our study also contributes to the situation vs. personality debate and show that abusive players tend to choose the leading role and players do not become more abusive when taking such roles.
We further demonstrate the predictive power of features based on team composition in prediction experiments.
\footnote{This is a long version of our WWW'19 short paper with additional analysis and prediction experiments: \url{https:/doi.org/10.1145/3308558.3312504}}
\end{abstract}

%% file: intro.tex
\section{Introduction}
\label{sec:intro}

\begin{figure}[t]\setlength{\abovecaptionskip}{0.cm}\setlength{\belowcaptionskip}{-0.2cm}
    \includegraphics[width=0.34\textwidth]{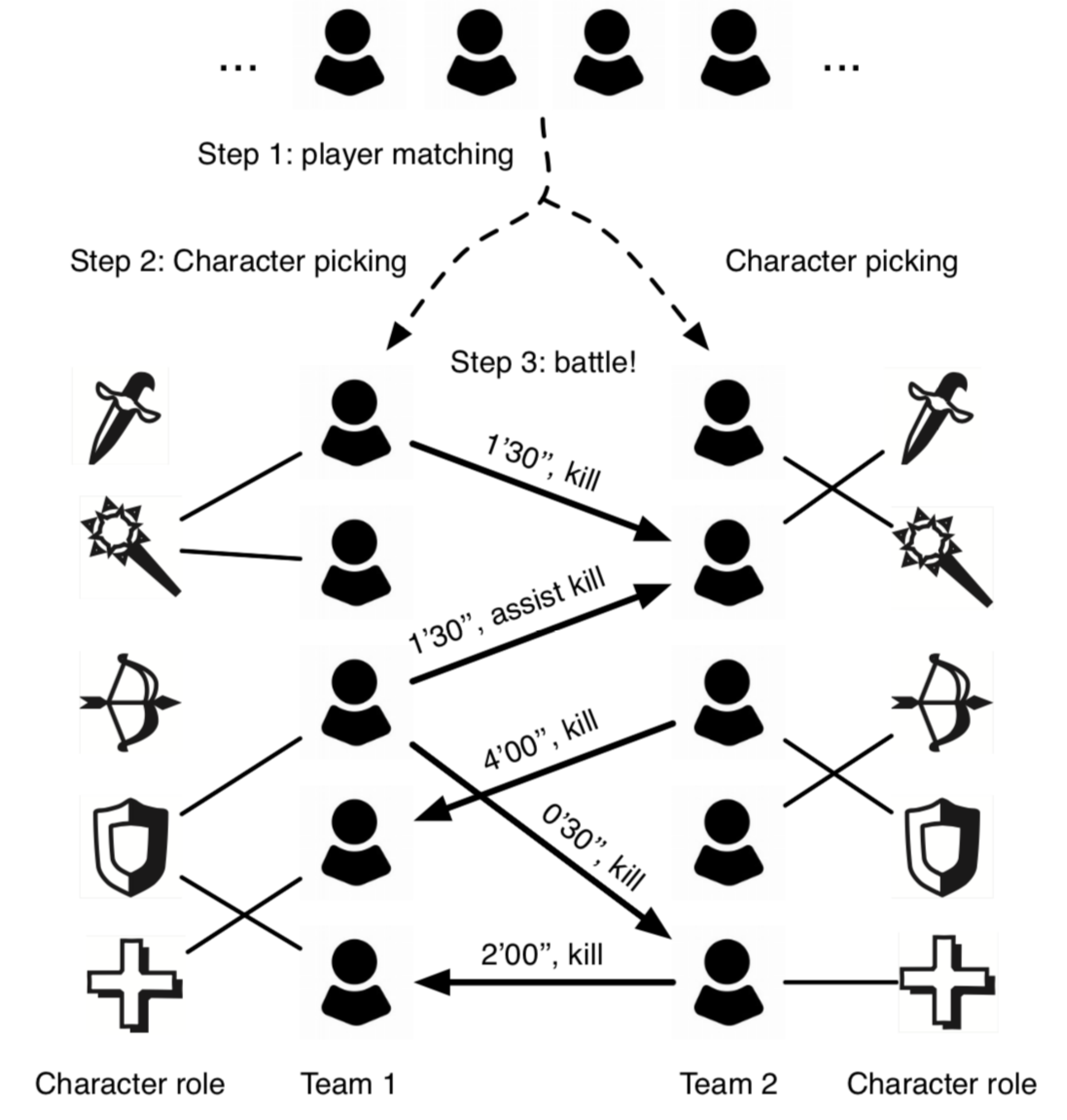}
    \caption{Illustration of games in Honor of Kings:
    1) a matching system groups 10 players into two similarly competitive teams;
    2) players in each team choose a character to control, which is designed to play a certain role (note that multiple players can choose the same role);
    3) two teams battle to destroy opponents' base.
    See the description of roles in Table~\ref{tb:role_character}. 
    }
    \label{fig:example}
\end{figure}

The increasing complexity and scale of tasks in modern society require individuals to work together as a team \cite{Wuchty:Science:2007,susan1997what}.
Crucial to the effectiveness of a team are the individuals in the team, i.e., team composition.
Extensive prior research has studied team roles to analyze team composition \cite{parker1990team,barbara1997,meredith2011management,partington1999team,stewart2005,belbin2012team}.
The most influential work is Belbin's team role framework, also known as Belbin Team Inventory.
A central hypothesis in Belbin's framework is that balance in team roles is associated with team performance.
\citet{belbin2012team} defines nine roles
\footnote{See \url{https://en.wikipedia.org/wiki/Team_Role_Inventories} for a quick explanation.}, 
and finds that teams with certain role combinations result in poor performance, even if they are formed by members with the sharpest mind and the most experience. 
However, team roles are usually implicit in real life and it is difficult to identify such nine roles in most contexts.

Another challenge in studying the effect of team composition on team effectiveness arises from the definition of {\em effectiveness}.
Team performance is the most straightforward definition and has been studied in a battery of studies \cite{salas2008teams,micheal1993understanding,susan1997what,dirks1999effects}.
Example measures of team performance include the impact of published papers from a team \cite{Wuchty:Science:2007}, winning a sports game \cite{JournalOfEconomicsAndBusiness:2002,Kim:CommunicationsOfTheAcm:}, etc.
However, the effectiveness of a team can also be reflected by the tenacity in face of adversity, and the rapport between team members \cite{bradner2005team}.
The effect of team composition likely depends on the definition of team effectiveness.
Therefore, it is important to understand the effect of team composition on the effectiveness of teams in multiple measures.

To address these two challenges, we identify multiplayer online battle arena (MOBA) as an ideal testbed and provide a large-scale study on the effect of team compositions on multiple measures of team effectiveness.
We use a dataset from \kingofglory, the largest MOBA game in the world,
and \figref{fig:example} illustrates the process of an example game.
This platform is ideal for studying the effect of team composition for three reasons.
First, there are defined roles in the game and players choose their roles to form a team.
Given that there are five players in each team, we are able to enumerate all possible team compositions.
We also have information about the characteristics of each role.
Second, the popularity of \kingofglory leads to digital traces of hundreds of millions of players in hundreds of millions of games.
These detailed game records allow us to explore the notion of team ``effectiveness'' beyond the simple measure of team performance, winning or losing.
For instance, we study the effect of team composition on abusive language use, which reflects the rapport in a team.
This research question naturally connects to the literature on 
toxic behavior in online communities and gaming \cite{cheng2017,Kwak2015,Cheng:ProceedingsOfIcwsm:2015}.

Finally, we have access to the past history of players for capturing their background and experience.
Historical information enables us to investigate questions in the ``situation vs. personality'' debate \cite{person-situation-09,Kenrick:AmericanPsychologist:1988}: how much of our behavior is
determined by fixed personality traits or by the specific situation at hand?
We will explore this question in the context of toxic behavior in online gaming.

\para{Organization and paper highlights.} In this paper, we conduct a large-scale study on the effect of team composition in the context of online gaming.
Our dataset comes from a popular MOBA game in China, \kingofglory, and includes 96 million games and 100 million players in total.
We provide details of the dataset and character roles in \secref{sec:data}.

We quantitatively study the effect of team composition on three distinct definitions of team effectiveness: 
(\secref{sec:performance}) team performance --- whether a team is going to win --- is directly based on the result/goal in a game and is the most common measure in prior studies;
(\secref{sec:surrender}) team tenacity --- whether a team is going to surrender --- indicates a team's resilience to adversity and is understudied in existing literature as most studies only look at individual tenacity or perseverance \cite{duckworth2007grit,baum2004relationship};
(\secref{sec:abuse}) toxic behavior, i.e., abusive language use, reflects the rapport in a team and is extracted from the activities during a game.

Our results on team effectiveness confirm the theory that balance in team roles is important for effective teams.
We find that diverse teams are more likely to win.
Furthermore, diverse teams are less likely to surrender, indicating a higher level of tenacity.
Note that the likelihood to surrender does not correlate with winning rate for commonly used team compositions.
However, we observe diverging effects on abusive language use:
diverse teams are more likely to abuse when losing, but are less likely to abuse when winning.

In addition to exploring the effect of team composition on team effectiveness, we investigate individual-level abusive language use to understand how teamwork influences individuals and shed light on the ``situation vs. personality'' debate \cite{person-situation-09,Kenrick:AmericanPsychologist:1988}.
We find that players who choose assassins (the leading role in a team) are more likely to abuse.
We further demonstrate that the reason is that abusive players are more likely to choose assassins, instead of the alternative explanation that a player becomes more abusive when playing assassins.
In a similar vein, we find that experienced players are more likely to abuse, but not when their teammates are also experienced.
A design implication is that matching players with similarly experienced players can improve the rapport in a team.

To demonstrate the effectiveness of modeling team composition, 
we formulate predictions tasks for all three measures of team effectiveness in \secref{sec:exp}.
We show that features based on team compositions can consistently outperform individual player information. 
In fact, modeling single roles and two-role combinations provides most of the predictive power.
Finally, we present related work in \secref{sec:related} and offer our concluding thoughts in \secref{sec:conclude}.

\begin{figure}[t]\setlength{\abovecaptionskip}{-0.cm}\setlength{\belowcaptionskip}{-0.cm}
	\includegraphics[width=0.25\textwidth]{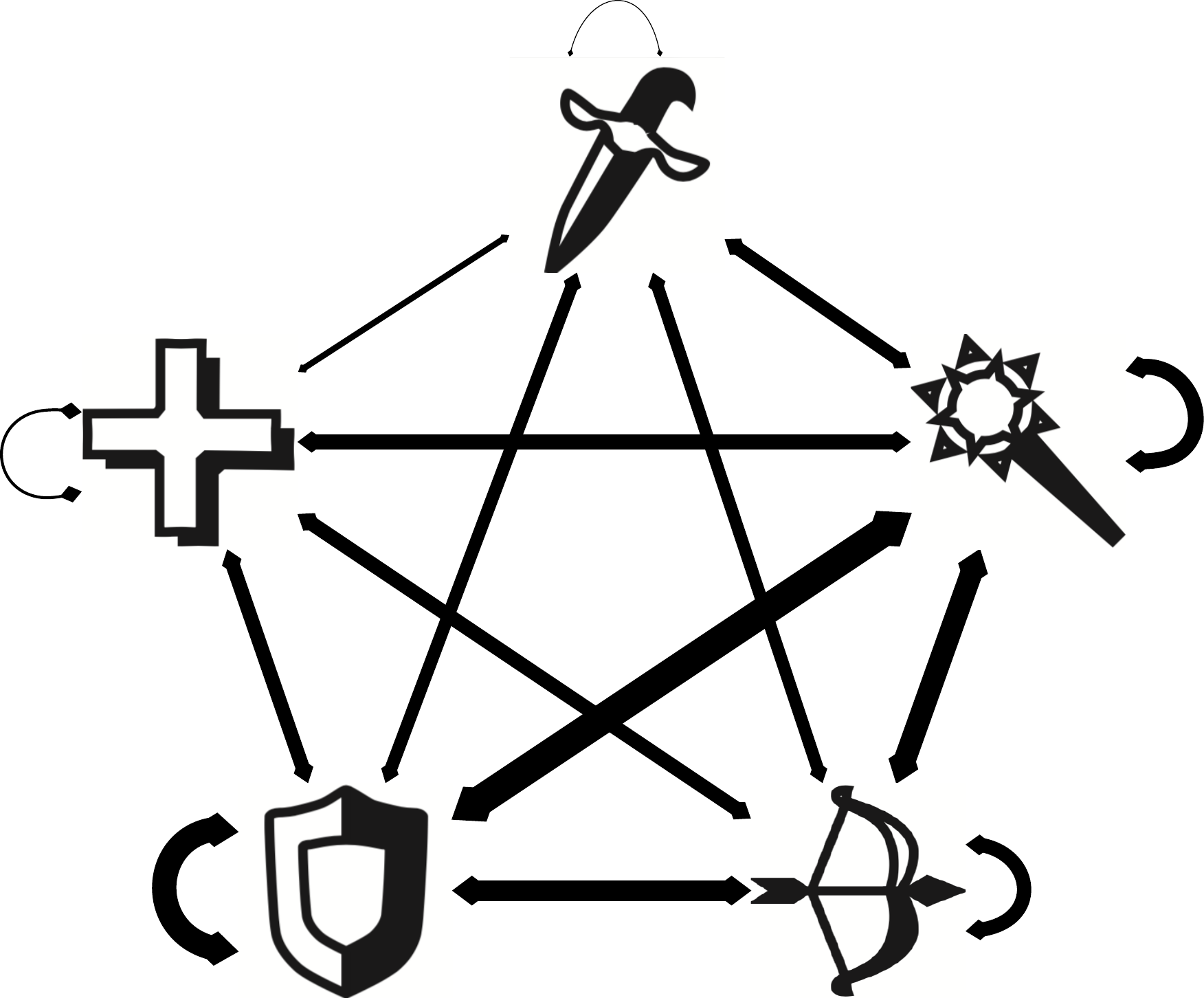}
	\caption{Role co-occurrence graph. Edges between two roles represent their co-occurrence in the same team. The width of edges is proportional to the square root of frequency. The three most common role pairs are warrior-warrior ($\protect\warrior$-$\protect\warrior$, 50.4\%), warrior-mage ($\protect\warrior$-$\protect\mage$, 47.8\%), and mage-marksman ($\protect\mage$-$\protect\marksman$, 43.8\%).
	}
	\label{fig:role_distribution}
\end{figure}

%% file: data.tex
\section{Dataset}
\label{sec:data}
Our dataset is provided by Tencent\footnote{\kingofglory is developed and published by Tencent, which is one of the largest Internet companies and also the largest game service provider in China. } and comes from a multiplayer online battle arena (MOBA) game: \kingofglory, 
the most profitable game in the world\footnote{See \url{https://en.wikipedia.org/wiki/Wangzhe_Rongyao} for an introduction of the native version of \kingofglory (Wangzhe Rongyao).}. 
Similar to many other MOBA games such as Dota2 and League of Legends, \kingofglory involves two teams to battle with each other.
Each team consists of five players and the success of a team depends on the chemistry and collaboration between the five team members.
The main motivation of our study is that players are expected to take certain roles when they select characters.
These roles are designed to complement each other and ``good'' teams usually require particular compositions of roles.
In this paper, we aim to explore the effect of team composition on different measures of team effectiveness.

There are multiple types of games in \kingofglory.
{\em Ranked games} are used to estimate a player's rank level in the game;
while the other games are for practice or for fun.
Since players care much less about unranked games than ranked games, all of our analyses are based on ranked games.
Unranked games are only used to analyze players' role preferences.

\begin{table}[t!]\setlength{\abovecaptionskip}{-0.2cm}\setlength{\belowcaptionskip}{-0.cm}
	\centering
	\small
	\begin{tabular}{l m{5.0cm}<{\raggedright} r}
		\hline
		Role & Description & Freq.(\%) \\ 
		\hline 
		warrior$\warrior$ & Warriors have a large pool of health and considerable damage; they often undertake the stress from enemies in side lanes. & 27.1\\
		\hline 
		mage$\mage$ & Mages have heavy spell damage, but a small pool of health; they often take the middle lane and prepare to assist teammates.& 25.8\\
		\hline 
		marksman$\marksman$ & 
		Marksmen have a small pool of heath but can cause heavy damage; 
		they need protection and may need time to grow.
		& 25.7\\
		\hline
		assassin$\assassin$ & 
		Assassins are explosive and control the pace, but the health pool is often small; they are usually the leading role in a team.
		& 11.5\\
		\hline 
		support$\support$ & Supports always aid teammates;
		they absorb damage while disrupting opponents by stunning and displacing them. & 9.9\\
		\hline
	\end{tabular}
	\normalsize
	\caption{Description and frequency of each role.}
	\label{tb:role_character}
\end{table}

\vpara{Dataset description.}
Our dataset is derived from daily logs spanning three weeks in the August of 2017.
Specifically, we first randomly sample 20K games in each day of the third week, thereby obtaining 140K games and 1.3M players participating in these games.
Among them, there are 78K ranked games. 
Our prediction experiments in \secref{sec:exp} are based on these ranked games, which we refer to as {\em prediction games}. 
We then retrieve all the games involving any of these 1.3M players in the first two weeks to obtain player information for prediction. 
In total, we obtain 95.6M games and 100.2M players. 
Among these games, there are 54.4M \textit{ranked games} on which most analyses in \secref{sec:performance}, \secref{sec:surrender}, and \secref{sec:abuse} are based, as under those analytical settings, we do not need historical information of every player in a team.
There are several exceptions where we need historical statistics of every team member, so we use prediction games in those cases and will point out in the text.

The log of each game includes its start time, end time, result (which team wins), gaming stats recorded when the game ends, and text messages in the chat room with abusive labels.
In addition, we have access to basic player information beyond our dataset, e.g., number of winning games, number of games played as MVP (Most Valuable Player) and total number of games in the past ten ranked games, etc.
See Table~\ref{tb:tag info} in \secref{sec:appendix} for details.

\vpara{Team roles.}
There are five roles in \kingofglory: 
warrior ($\warrior$), mage ($\mage$), marksman ($\marksman$), assassin ($\assassin$), and support ($\support$).
A team is composed of different roles depending on each player's choice.
Henceforth, we use five icons to represent a team composition.
For instance, $(\warrior,\mage,\mage,\assassin,\assassin)$ stands for a team with one warrior, two mages, two assassins, and no marksman or support. 
In total, there exist 126 team compositions. 
Table~\ref{tb:role_character} presents the characteristics and frequency of each role\footnote{The boundary between roles can sometimes be blurry, especially considering different player tactics.}.
Figure~\ref{fig:role_distribution} shows the frequency of two roles co-occurring in the same team and \tableref{tbl:roles_winrate} shows several role combinations with its winning rate and frequency.
We can clearly see that team compositions are not used equally by players:
single roles and pairs of roles both vary greatly in frequency. 
Please refer \tableref{tb:basic statistics} for detailed statistics of our dataset.

%% file: performance.tex
\section{Team Composition and Winning}
\label{sec:performance}

We first investigate the effect of team composition on whether a team wins or not, a direct measure of team performance. 
Although the matching system in online gaming is designed to balance the ability of two teams, we observe that the winning rate of most team compositions is significantly lower than 50\%, indicating that some role combinations cannot effectively work with each other.

\begin{table}[tp!]\setlength{\abovecaptionskip}{-0.1cm}\setlength{\belowcaptionskip}{-0.cm}
	\centering
	\begin{tabular}{p{4.2cm} p{3cm}}
		\toprule
		Statistics & Number(fraction/SEM) \\
		\toprule
		\multicolumn{2}{c}{Overall statistics}\\
		$\#$ranked games & $54,434,817$\\
		$\#$team compositions & $126$\\
		$\#$surrender games & $5,807,005~(10.7\%)$\\
		$\#$abusing games& $30,668,577~(56.3\%)$\\
		$\#$early games($t \le$ 11 min)& $5,049,934~(9.3\%)$\\
		$\#$middle games(11 < $t\le$ 16 min)& $23,089,993~(42.4\%)$\\
		$\#$late games($t$ > 16 min)& $26,294,890~(48.3\%)$\\
		\hline
		\multicolumn{2}{c}{Top five frequent combinations}\\
		$\assassin-\mage-\support-\marksman-\warrior$ & $16,186,003~(14.9\%)$\\
		$\mage-\support-\marksman-\warrior-\warrior$ & $14,516,443~(13.3\%)$\\
		$\assassin-\mage-\marksman-\warrior-\warrior$ & $12,630,805~(11.6\%)$ \\
		$\mage-\marksman-\warrior-\warrior-\warrior$ & $9,124,177~(8.4\%)$ \\
		$\mage-\mage-\marksman-\warrior-\warrior$ & $8,929,680~(8.2\%)$ \\
		\hline
		\multicolumn{2}{c}{Personal statistics}\\
		$\#$target players & $1,306,754$\\
		$\#$target players who have played ranked games& $1,173,372$\\
		$\#$ranked games each player has & $48.4~(0.05)$\\
		average winrate & $52.4\%~(1.0\times10^{-4})$\\
		average abusing probability & $9.3\%~(1.0\times10^{-4})$\\
		average surrender probability & $5.0\%~(6.9\times10^{-5})$\\
		$\#$target players who have played at least 20 ranked games & $758,494~(58.0\%)$\\
		\bottomrule
	\end{tabular}
	\caption{Statistics of the dataset.}
	\label{tb:basic statistics}
\end{table}

\begin{figure*}[t]\setlength{\abovecaptionskip}{-0.0cm}\setlength{\belowcaptionskip}{-0.cm}
	\begin{subfigure}[t]{0.256\textwidth}
		\includegraphics[width=\textwidth]{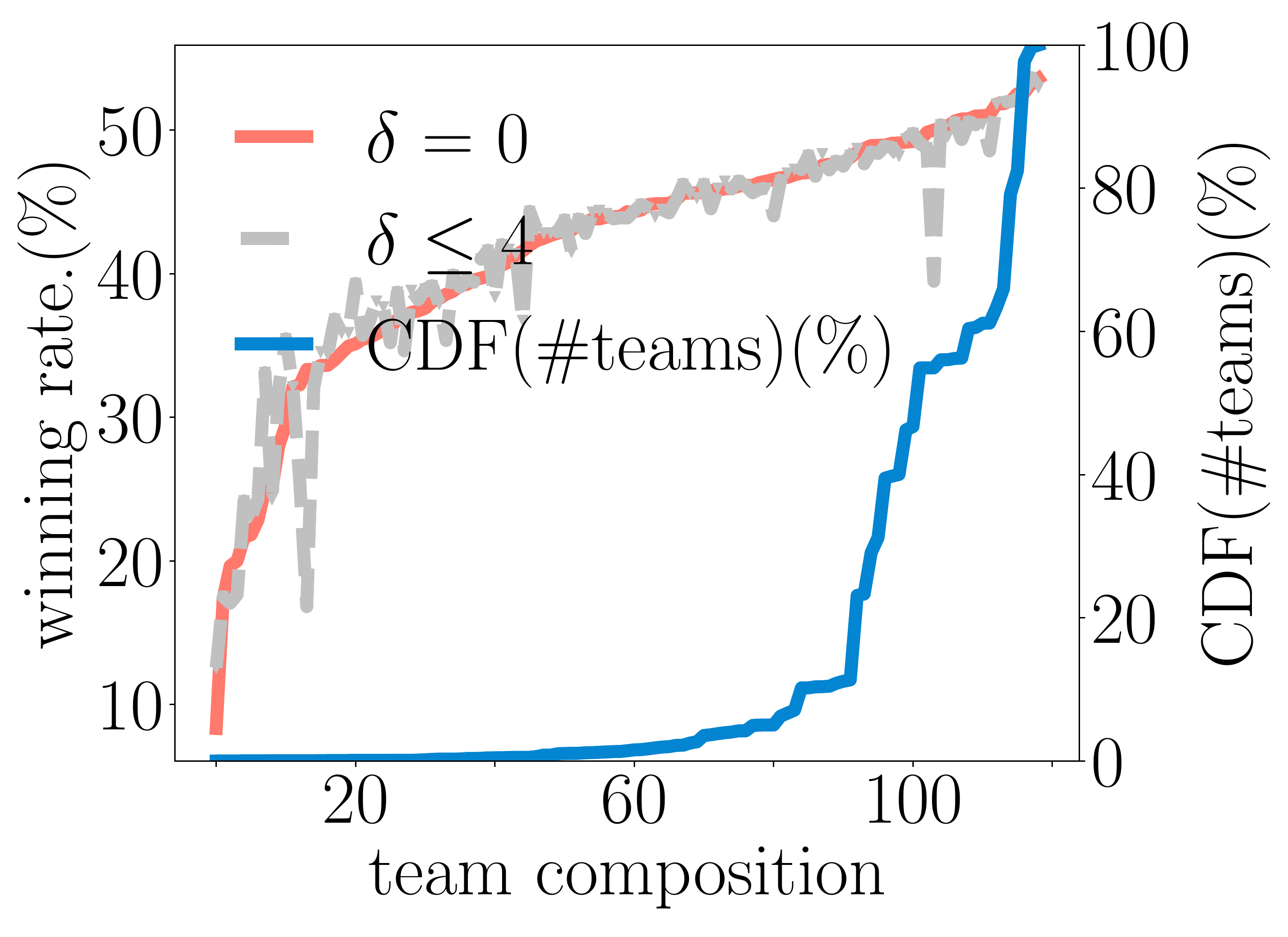}
		\caption{Sorted winning rate of different team compositions. 
}
		\label{fig:sorted_role_winrate}
	\end{subfigure}
	\hfill
	\begin{subfigure}[t]{0.23\textwidth}
		\includegraphics[width=\textwidth]{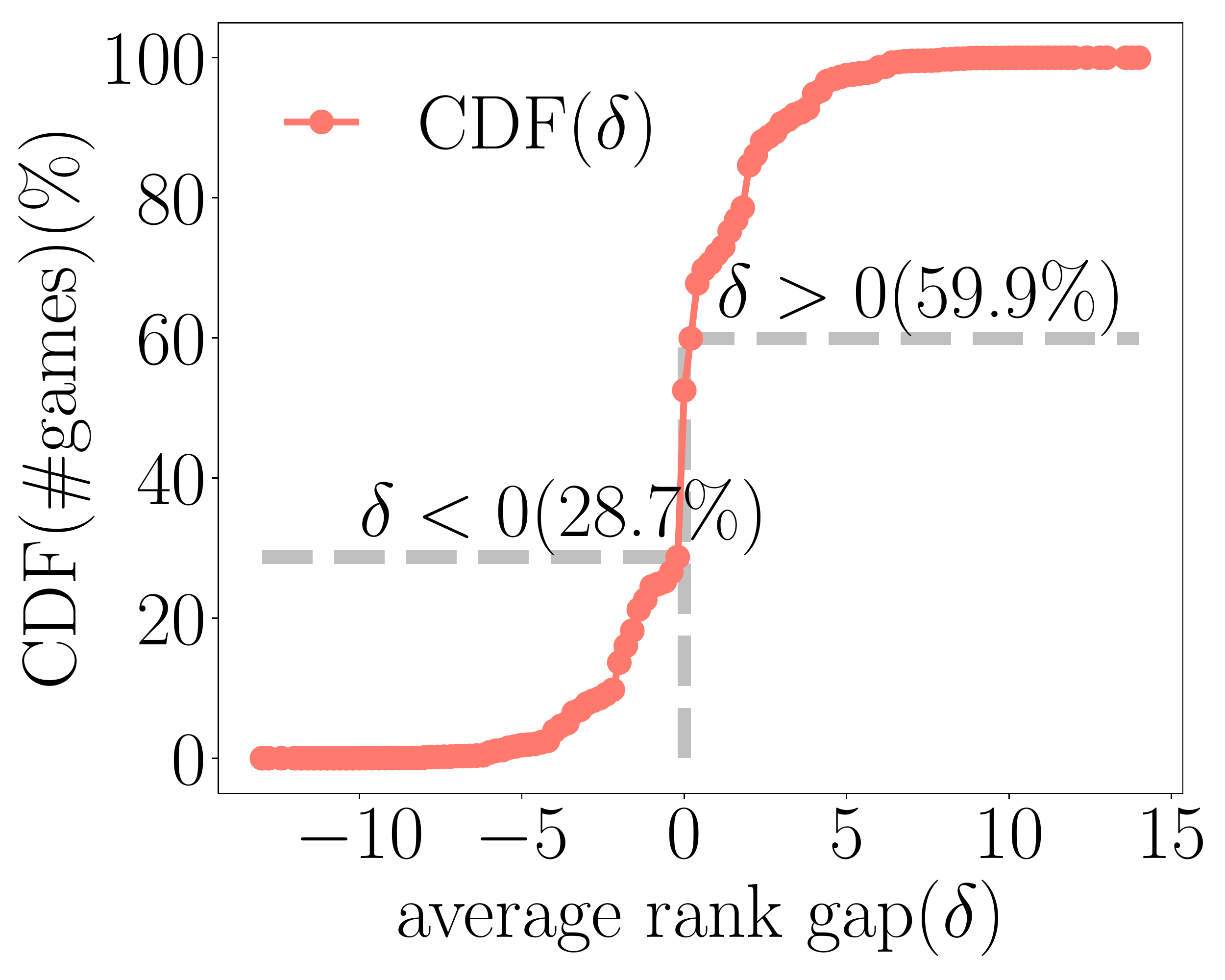}
		\caption{Distribution of rank gap ($\delta$) between two teams.}
		\label{fig:rankgap distribution}
	\end{subfigure}
	\hfill
	\begin{subfigure}[t]{0.23\textwidth}
		\includegraphics[width=\textwidth]{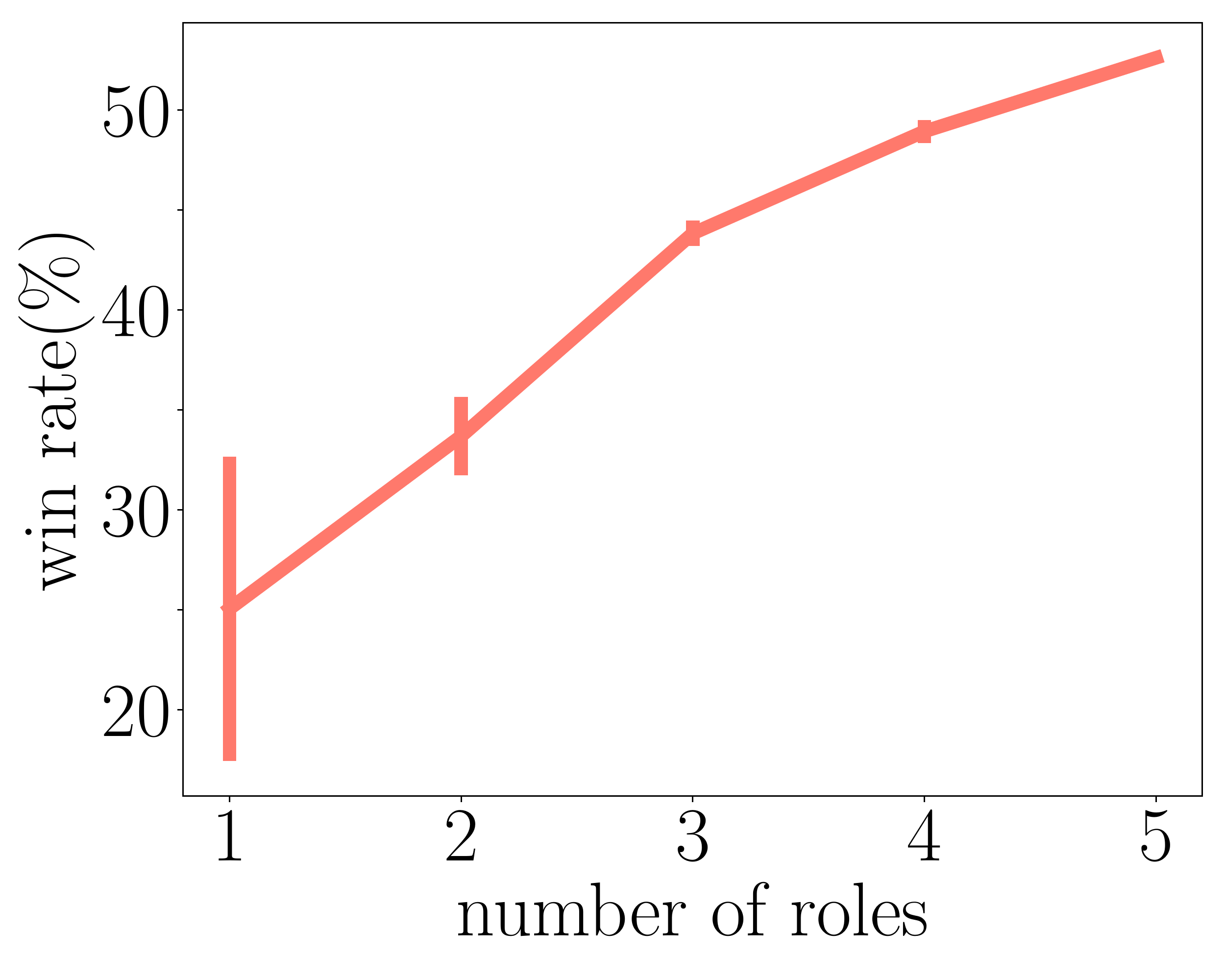}
		\caption{Winning rate vs. team diversity (rank gap $\delta=0$).}
		\label{fig:roles_winrate_group}
	\end{subfigure}
	\hfill
	\begin{subfigure}[t]{0.23\textwidth}
		\includegraphics[width=\textwidth]{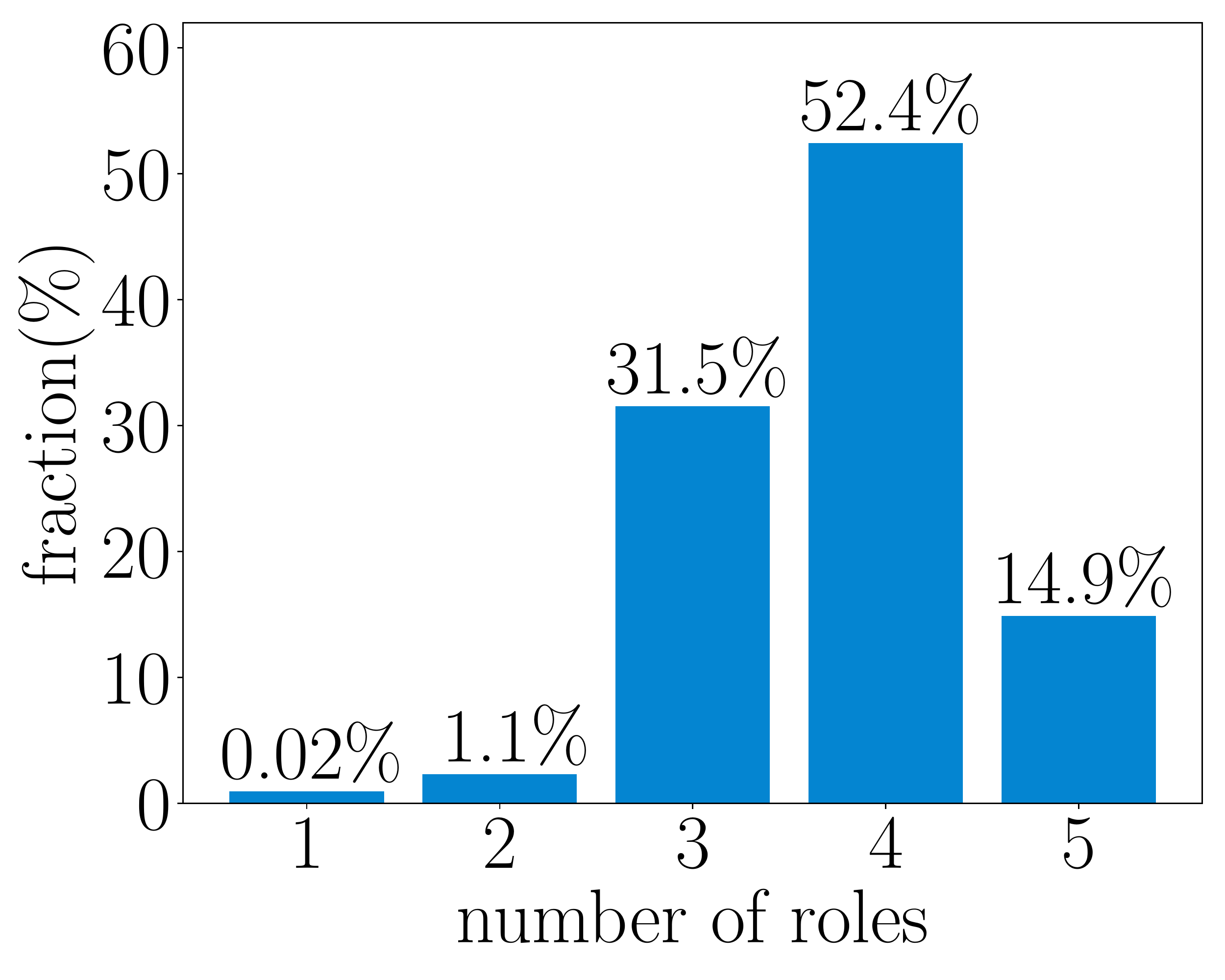}
		\caption{Frequency of teams vs. team diversity.}
		\label{fig:role_dis}
	\end{subfigure}
	\caption{Team composition and winning rates. In \figref{fig:sorted_role_winrate},
		$x$-value corresponds to the index of a team composition, 
		and is sorted by winning rate.
		The CDF of \#teams shows the cumulative distribution in \#teams according to the order in the $x$-axis.
		\figref{fig:rankgap distribution} shows the distribution of the rank gap between two teams. 
		\figref{fig:roles_winrate_group} shows that a diverse team with more role categories is more likely to win (error bars represent standard errors). 
		\figref{fig:role_dis} shows that most teams are rationale and cover at least 3 roles.
	}
	\label{fig:composition vs. win}
\end{figure*}

\vpara{Although two teams are with similar ``skills'', there exist winning and losing team compositions (\figref{fig:sorted_role_winrate}). }
Since we have 126 different team compositions in total, it is straightforward to estimate the {\em winning rate} of each team composition using the fraction of winning games, $\frac{\#\text{winning~games}}{\#\text{games}}$.
\figref{fig:sorted_role_winrate} shows that the winning rate of different team compositions spans a wide range, from 8.3\% to 53.6\%. 
It seems that some team compositions are doomed to lose. 
In addition, the figure also shows the cumulative distribution of \#teams for each team composition, which suggests that most losing teams do not occur frequently.
Notice that a similar plot will also be made for surrendering and abusing and we will see that the CDFs present very different shapes.

\begin{table}[ht]\setlength{\abovecaptionskip}{-0.1cm}\setlength{\belowcaptionskip}{-0.cm}
	\centering
	\small
	\begin{tabular}{p{4cm}rr}
		\toprule
		Team compositions & Win rate & Used frequency(\%) \\
		\midrule
		\multicolumn{3}{c}{team compositions with the highest winning rate} \\
		$\mage-\support-\support-\marksman-\marksman$ & 53.6\% & 0.3\\
		$\assassin-\mage-\support-\support-\marksman$ & 53.2\% & 2.0\\
		$\assassin-\mage-\support-\marksman-\warrior$ & 52.6\% & 14.9\\
		\midrule
		\multicolumn{3}{c}{team compositions with the lowest winning rate} \\
		$\mage-\mage-\support-\support-\support$  & 19.7\% & $3.1 \times 10^{-3}$\\
		$\assassin-\assassin-\assassin-\assassin-\assassin$ & 17.4\% & $2.2 \times 10^{-2}$\\
		$\mage-\support-\support-\support-\support$ & 8.3\% & $7.0 \times 10^{-4}$\\
		\midrule
		\multicolumn{3}{c}{team compositions that are most used} \\
		$\assassin-\mage-\support-\marksman-\warrior$ & 52.6\% & 14.9\\
		$\mage-\support-\marksman-\warrior-\warrior$ & 52.0\% & 13.3\\
		$\assassin-\mage-\marksman-\warrior-\warrior$ & 48.5\% & 11.6\\
		\bottomrule
	\end{tabular}
	\caption{Example team compositions and winning rates.
	}
	\label{tbl:roles_winrate}
\end{table}

The fluctuation of winning rate between different team compositions among all games may not due to the effect of team compositions;
another possible explanation is the ability of individual players: only relatively bad players choose certain bad team compositions.
Although a matching system is designed to balance the ability of the players in two teams,
it is difficult to always make sure that two teams exactly have the same ability.
\figref{fig:rankgap distribution}
shows the distribution of the rank level gap ($\delta$) between two teams\footnote{
Rank level indicates a player's gaming skill and ranges from 0 to 26.} . 
To calculate $\delta$, we sum up the individual rank level in a team, and subtract the result of losing team from that of winning team. 
By the effect of the matching system, the rank gaps in over $90\%$ games are relatively small (within $\pm 5$). 
These minor gaps, however, are still associated with a significant difference in winning rate: 
only $28.7\%$ games end up with the winning of the lower ranked team.
Therefore, to demonstrate the influence of team compositions on winning rates and to exclude the ability factor, we control the rank gap between teams in \figref{fig:sorted_role_winrate}: 
we focus on games with $\delta = 0$ (red plots) and $\delta \leq 4$ (gray plots) respectively. 

To further illustrate winning and losing team compositions, we present team compositions with the highest (lowest) winning rate in Table~\ref{tbl:roles_winrate}\footnote{
	We make similar tables as \tableref{tbl:roles_winrate} for surrendering and abusing.
	See \secref{sec:appendix} for reference.} (controlling the rank gap $\delta=0$). 
Upon a careful examination, these losing teams do not often occur, reflected also by the cumulative distribution curve in \figref{fig:sorted_role_winrate}.
In total, the bottom $90$ losing team compositions only take $11.1\%$ of all teams.
This suggests that most players choose a reasonable team composition. 
However, the last three rows in Table~\ref{tbl:roles_winrate} show that the most common team compositions still vary in winning rates, from 48.5\% to 52.6\%.

\begin{figure*}[t]\setlength{\abovecaptionskip}{-0.0cm}\setlength{\belowcaptionskip}{-0.cm}
	\begin{subfigure}[t]{0.32\textwidth}
		\centering
		\includegraphics[width=0.78\textwidth]{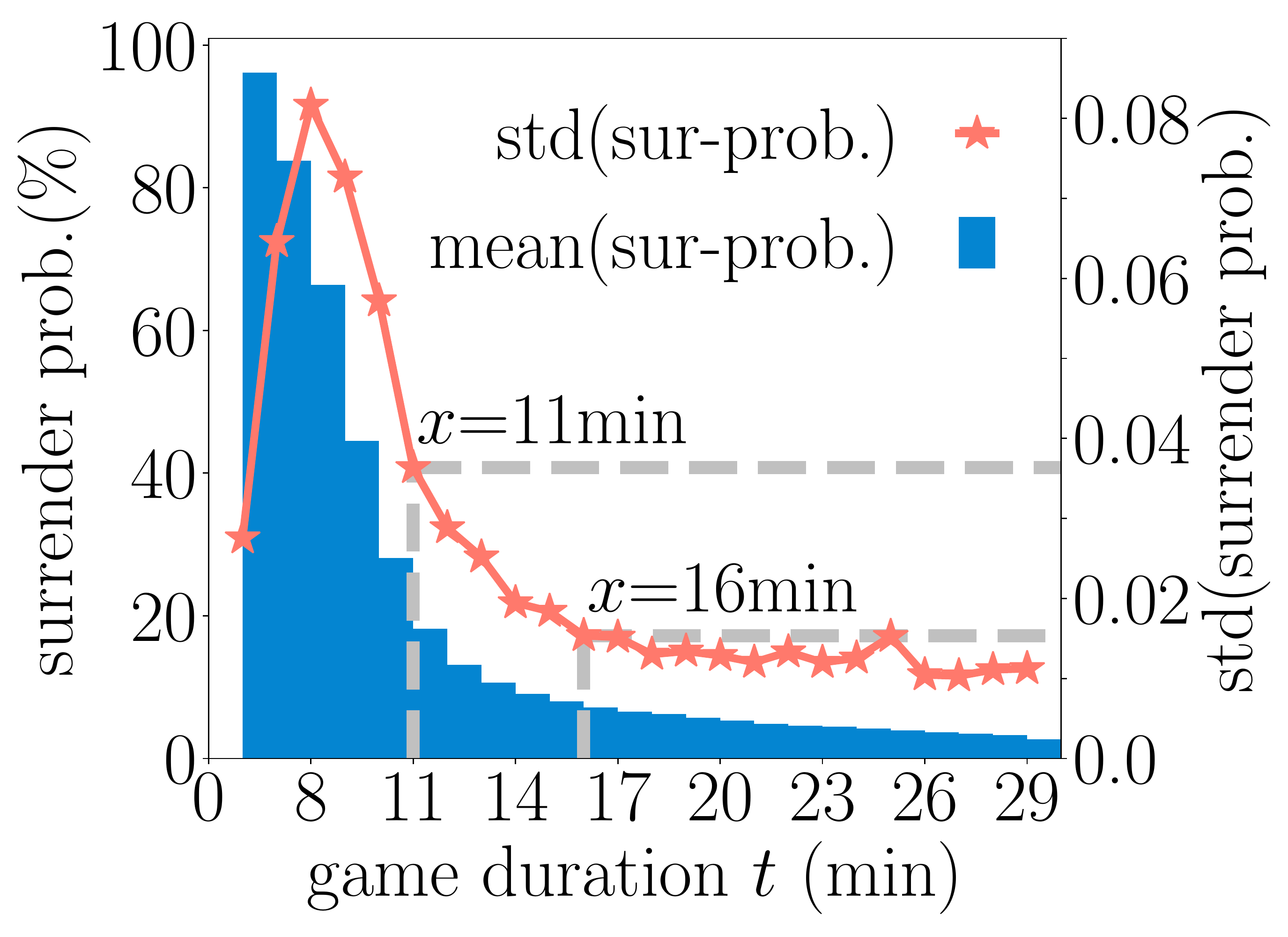}
		\caption{Surrender probability vs. game duration. 
		}
		\label{fig:surrender:time}
	\end{subfigure}
	\begin{subfigure}[t]{0.32\textwidth}
		\centering
		\includegraphics[width=0.78\textwidth]{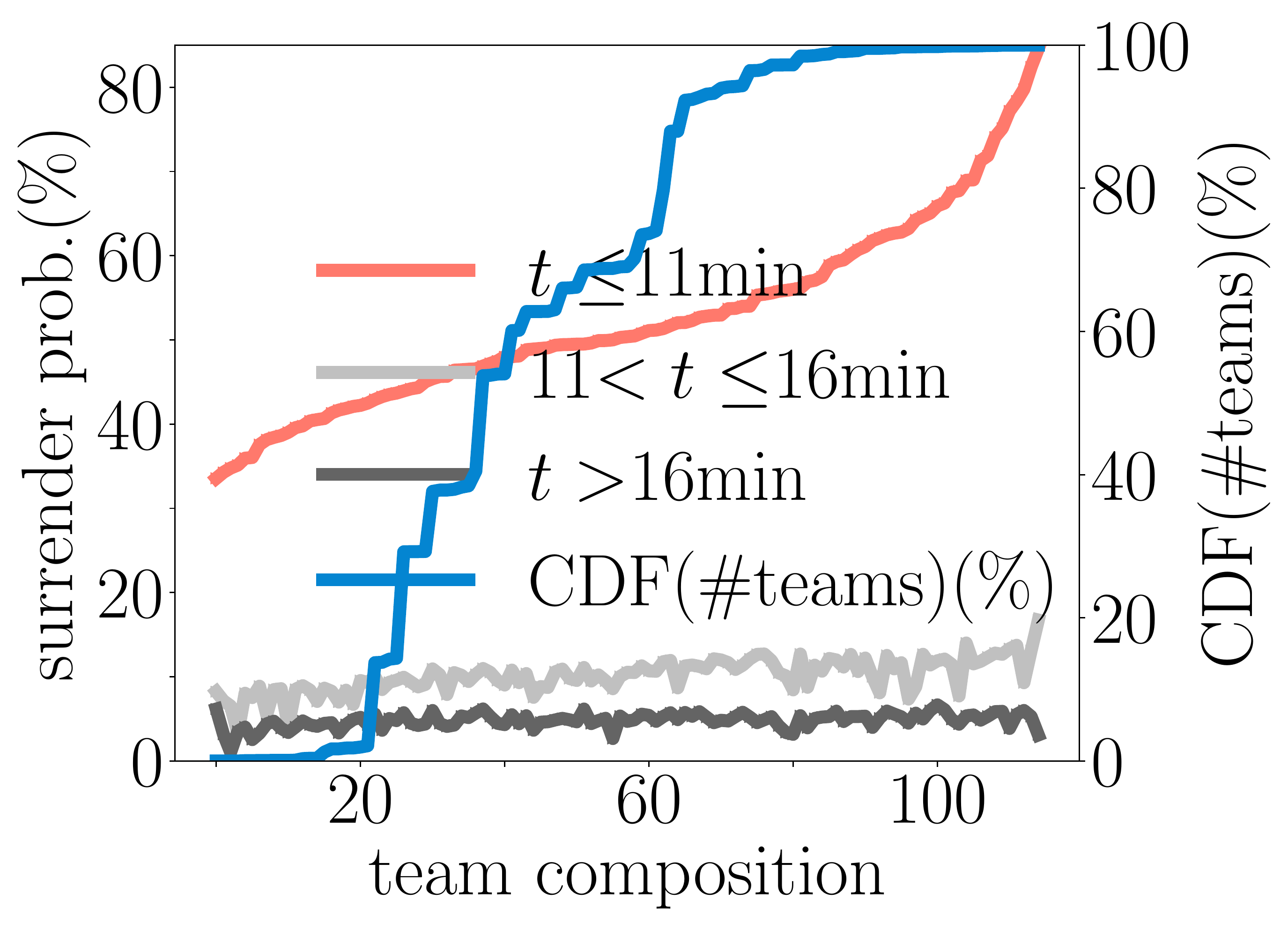}
		\caption{Surrender prob. vs. team composition.
		}
		\label{fig:surrender:sorted}
	\end{subfigure}
	\begin{subfigure}[t]{0.32\textwidth}
		\centering
		\includegraphics[width=0.7\textwidth]{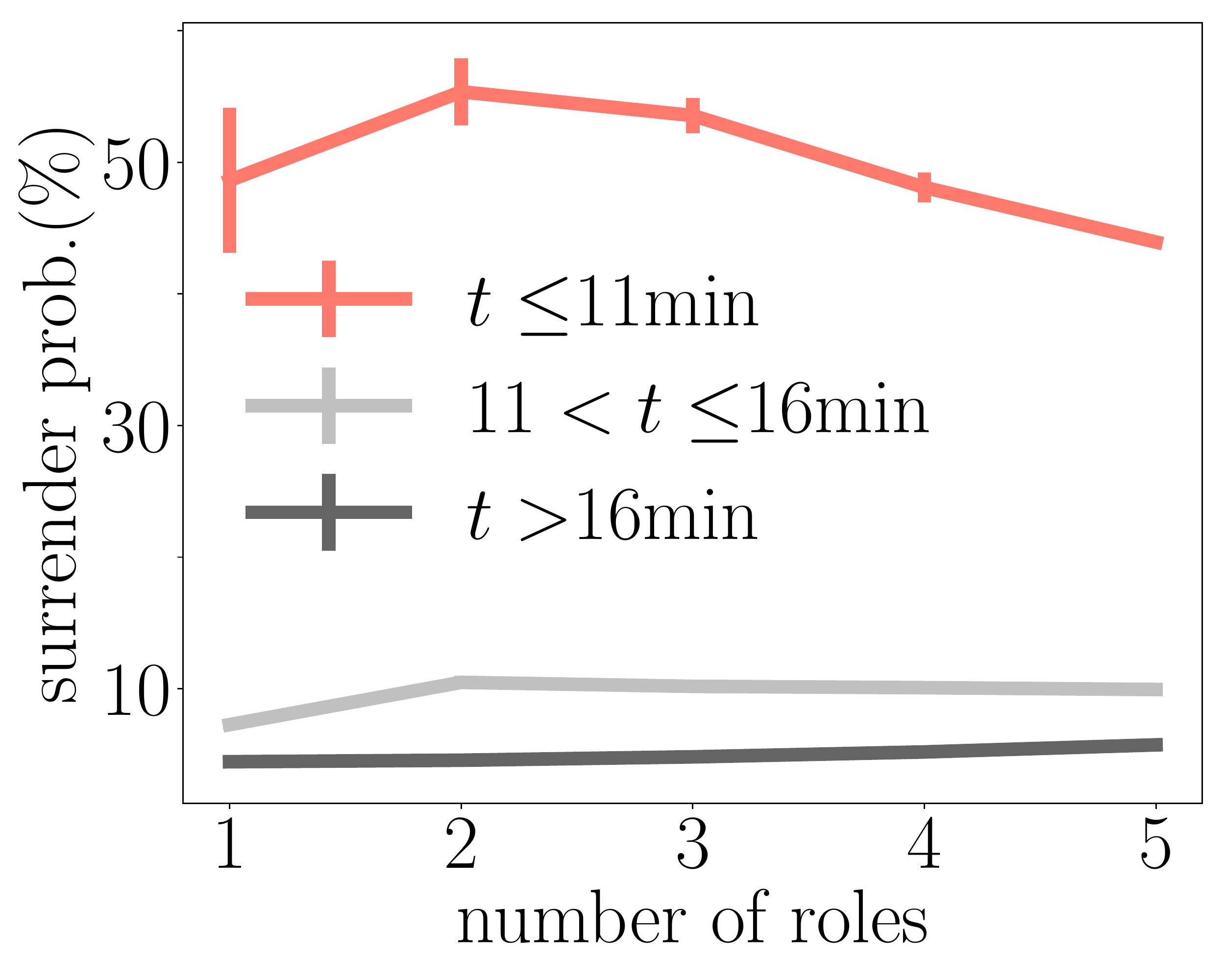}
		\caption{Surrender prob. vs. team diversity.}
		\label{fig:roles variety(surrender)}
	\end{subfigure}
	\caption{
		Team composition and surrender probability. 
		\figref{fig:surrender:time} shows the mean and standard deviation of surrender probability across team compositions conditioned on game duration. 
		\figref{fig:surrender:sorted} shows that team compositions have varying surrender probabilities in games that end within 11 minutes, but have similar ones in longer games.
		\figref{fig:roles variety(surrender)} shows that diverse teams are less likely to surrender in early games (error bars represent standard errors).
		\label{fig:surrender:composition}
	}
\end{figure*}

\vpara{Diverse teams tend to perform well (\figref{fig:roles_winrate_group}, \ref{fig:role_dis}).}
An observation that stands out in Table~\ref{tbl:roles_winrate} is that teams with the highest winning rate consist of more role categories than teams with the lowest winning rate.
For instance, support ($\support$) or assassin($\assassin$) dominates the two compositions with the lowest winning rate, while all the top winning teams have at least four roles.

We further explore this observation by examining how winning rate changes as the number of roles in a team grows (\figref{fig:roles_winrate_group}). 
We find that diverse team compositions are much more likely to win.
One explanation is that these roles are designed to complement each other and teams with too few roles have weaknesses to be exploited. 
For instance, a team with no mage ($\mage$) or marksman ($\marksman$) cannot cause sufficient damage in a game. 
\figref{fig:role_dis} shows that most teams are employing at least three role categories, which is consistent with the observation in \figref{fig:sorted_role_winrate} that the losing team compositions on the left are not usually used.

%% file: surrender.tex
\section{Team Composition and Surrendering}
\label{sec:surrender}
As nothing worth achieving is going to be easy,
tenacity is an important characteristic of an effective team.
In the context of \kingofglory, team tenacity can be reflected by not surrendering easily. 
The surrendering procedure is as follows: a player can propose to surrender and if at least four of five team members agree, the team will surrender as a whole and lose immediately. 
In this section, we examine the effect of team composition on team tenacity.

\para{Surrender probability varies across team compositions in early games (\figref{fig:surrender:time}, \ref{fig:surrender:sorted}).}
Similar to winning, surrender happens at the end of a game.
But surrender requires extra careful analysis because
surrender may not always reflect team tenacity in \kingofglory and other MOBA games.
For example, when the opponents are destroying the team base and all team members are killed, it is time to move on and surrender is simply a sign of ``game over''. 
This is common in late games. 
Therefore, it is necessary to distinguish surrender in different game stages. 

We examine the mean and the standard deviation of surrender probability across team compositions conditioned on game duration in \figref{fig:surrender:time}.
We make two observations.
First, the surrender probability monotonically decreases as game duration increases, partly because a team is only allowed to surrender starting from the 6th minute and games end in 6-8 minutes mostly because a team surrenders.
Second, the standard deviation across team compositions first increases and then decreases as game duration increases, indicating that in late games every team surrenders with similar probability to signal game over.
Based on \figref{fig:surrender:time}, we define three game stages: 
\textit{early games ($t \leq 11$)}, where the surrender probability has great mean and great std; 
\textit{late games ($t > 16$)}, where the surrender probability has small mean and small std;
\textit{middle games ($11 < t \leq 16$)}, the middle part\footnote{A gaming related reason for this split is that 11-minutes marks a watershed because the ``Overlord'' and ``Dark Tyrant'' (two important non-player characters) appear at the 10$^{\mbox{th}}$ minute and game state can change significantly from 11 to 16 minutes when the characters get killed.}. 
Proportions of games in each stages are shown in \tableref{tb:basic statistics}.

Having established the three types of games, we hypothesize that surrender in early games is the most indicative of team tenacity and varies across team compositions, while surrender in middle and late games is more of a formality and should not depend on team compositions.
To study that,
we sort team compositions by surrender probability in early games in \figref{fig:surrender:sorted}.
According to the same order, we show the surrender probability in middle games and late games, as well as the cumulative distribution function of \#teams.
Consistent with our hypothesis,
the surrender probability of different team compositions span a wide range from 33.6\% to 84.6\% in early games, but is pretty stable in middle games and late games.
The CDF of \#teams presents a different shape from that in \figref{fig:sorted_role_winrate}: commonly used team compositions are neither the most tenacious nor the least tenacious.
\tableref{tbl:roles-surrender} shows the most tenacious and the least tenacious team compositions.
 
\vpara{Diverse teams tend to be tenacious (\figref{fig:roles variety(surrender)}).} 
Role diversity influences surrender probability in early games, especially when the number of roles is large.
It seems that diverse teams with five roles are the most tenacious in early games.
In comparison, the influence on middle and late games is minimal. 

\begin{figure}[t]\setlength{\abovecaptionskip}{-0.0cm}\setlength{\belowcaptionskip}{-0.cm}
	\begin{subfigure}[t]{0.23\textwidth}
		\includegraphics[width=\textwidth]{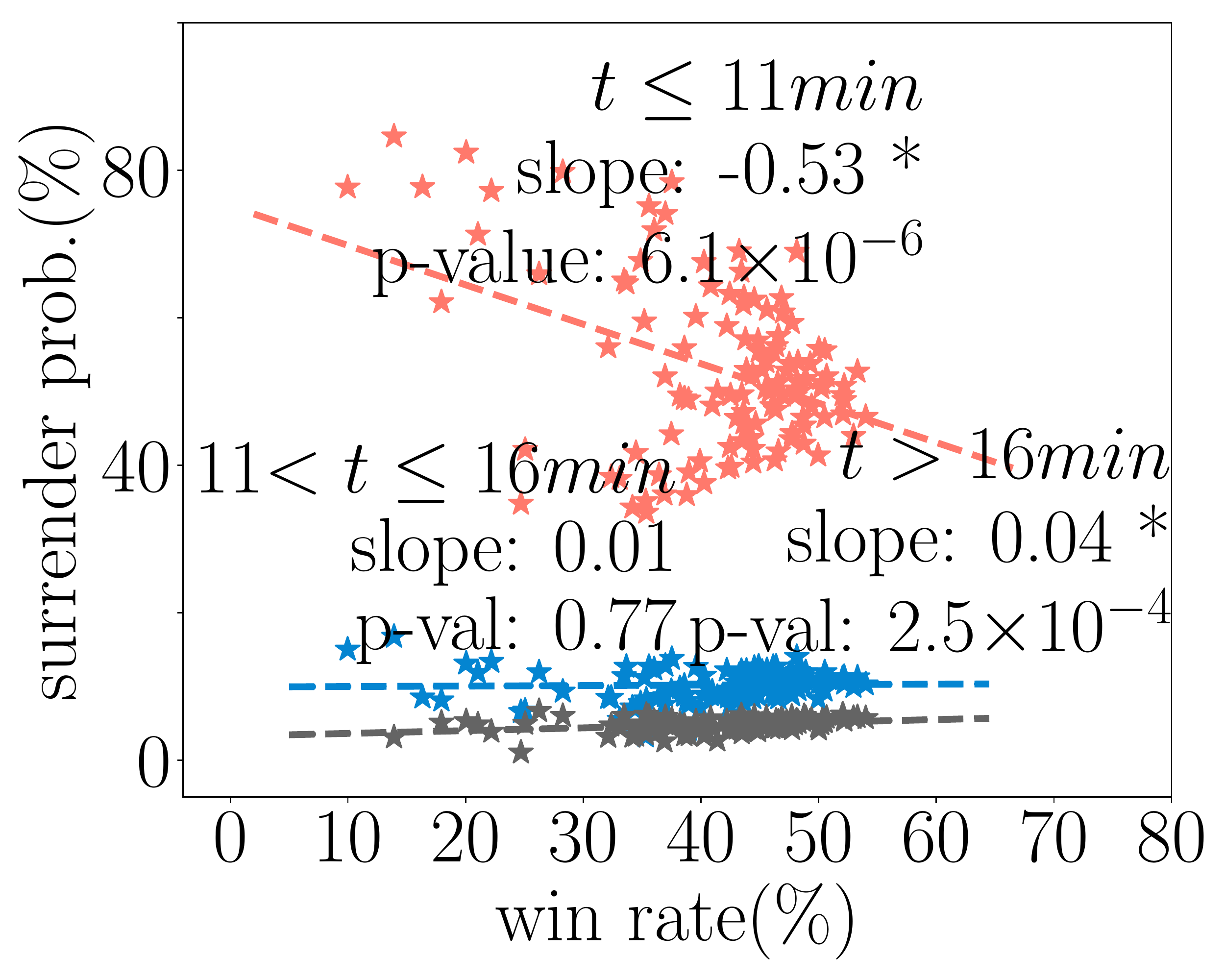}
		\caption{All team compositions.}
		\label{fig:surrender vs winrate}
	\end{subfigure}
	\begin{subfigure}[t]{0.23\textwidth}
		\includegraphics[width=\textwidth]{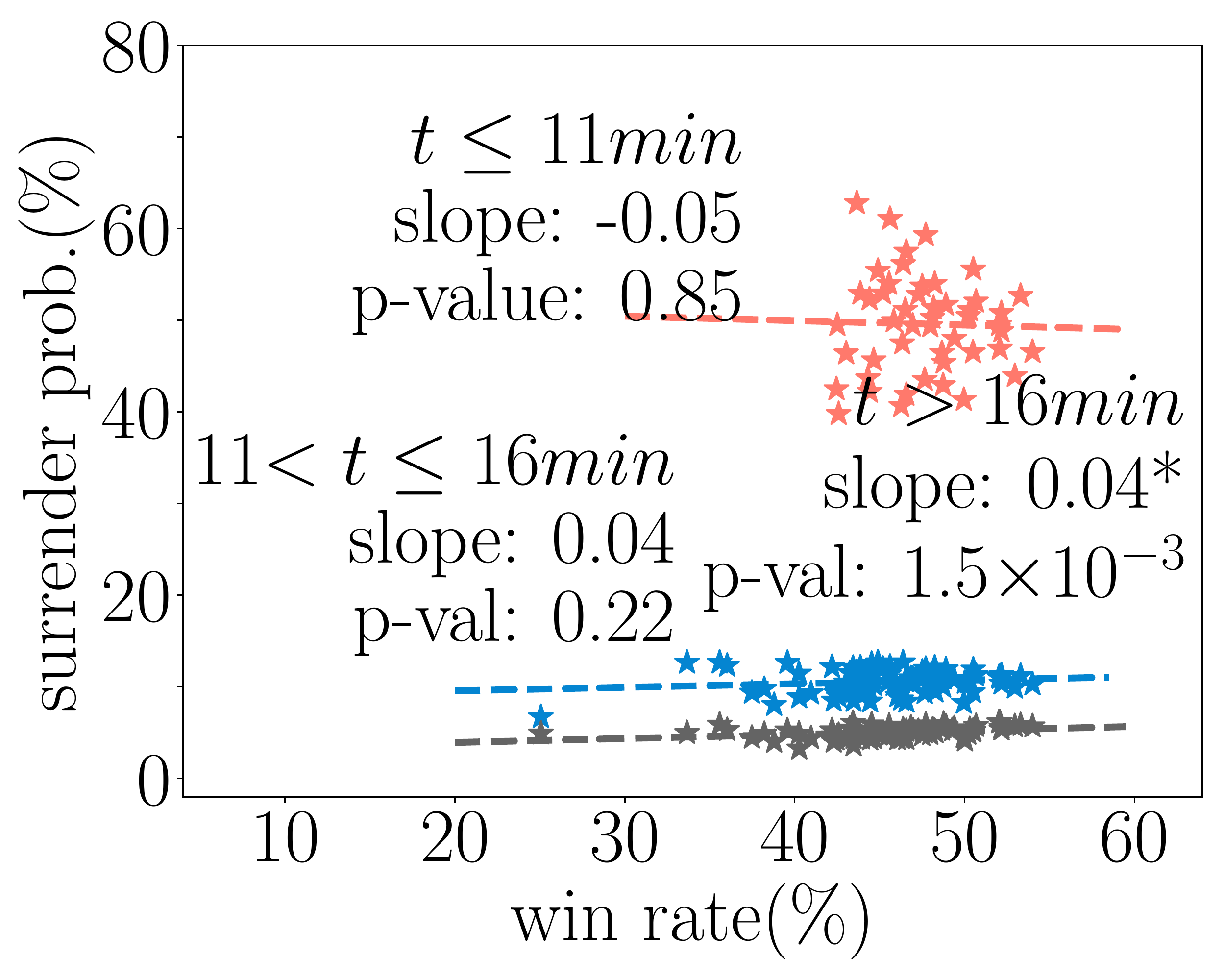}
		\caption{Filtering infrequent team compositions.}
		\label{fig:surrender vs winrate threshold}
	\end{subfigure}
	\caption{Surrender probability vs. winning rate.
		In both figures, each dot denotes a team composition with its $x$-value for winning rate and $y$-value for surrender probability.
		\figref{fig:surrender vs winrate} shows results for all team compositions.  \figref{fig:surrender vs winrate threshold} shows the results after removing infrequent compositions, 
		where surrender is not correlated with winning rate.
	}
	\label{fig:surrender:vs winrate}
	\vspace{-10pt}
\end{figure}

\para{A weak team can still be tenacious (\figref{fig:surrender:vs winrate}).} 
We have shown that diverse teams tend to both perform well and show tenacity in adversity.
One concern is that these two measures are correlated and tenacity is simply a side effect of team strength (winning).
This hypothesis suggests that winning team compositions should be less likely to surrender.
To further understand this issue, we study the correlation between surrender probability and winning rate for different team compositions.
\figref{fig:surrender vs winrate} shows that there is indeed a negative correlation between winning rate and surrender probability in early games.
There is little correlation in middle and late games.\footnote{In fact, there is a small significantly positive coefficient in late games, which is another evidence that strength does not entail tenacity.}
However,
it seems that the correlation between winning rate and surrender probability in early games is dominated by a few outliers with very low winning rates. 
Given that we know that these teams only take a small fraction of all games in \secref{sec:performance}, 
we filter infrequent team compositions that appear less than 10,000 times and find that in the remaining team compositions ($98.7\%$ of all teams), 
surrender probability is not correlated with winning rate in all game stages (\figref{fig:surrender vs winrate threshold}), suggesting that team tenacity is almost independent of team strength/performance. 

%% file: abuse.tex
\begin{figure*}[t]\setlength{\abovecaptionskip}{-0.05cm}\setlength{\belowcaptionskip}{-0.cm}
	\centering
	\begin{subfigure}[t]{0.32\textwidth}
		\centering
		\includegraphics[width=0.78\textwidth]{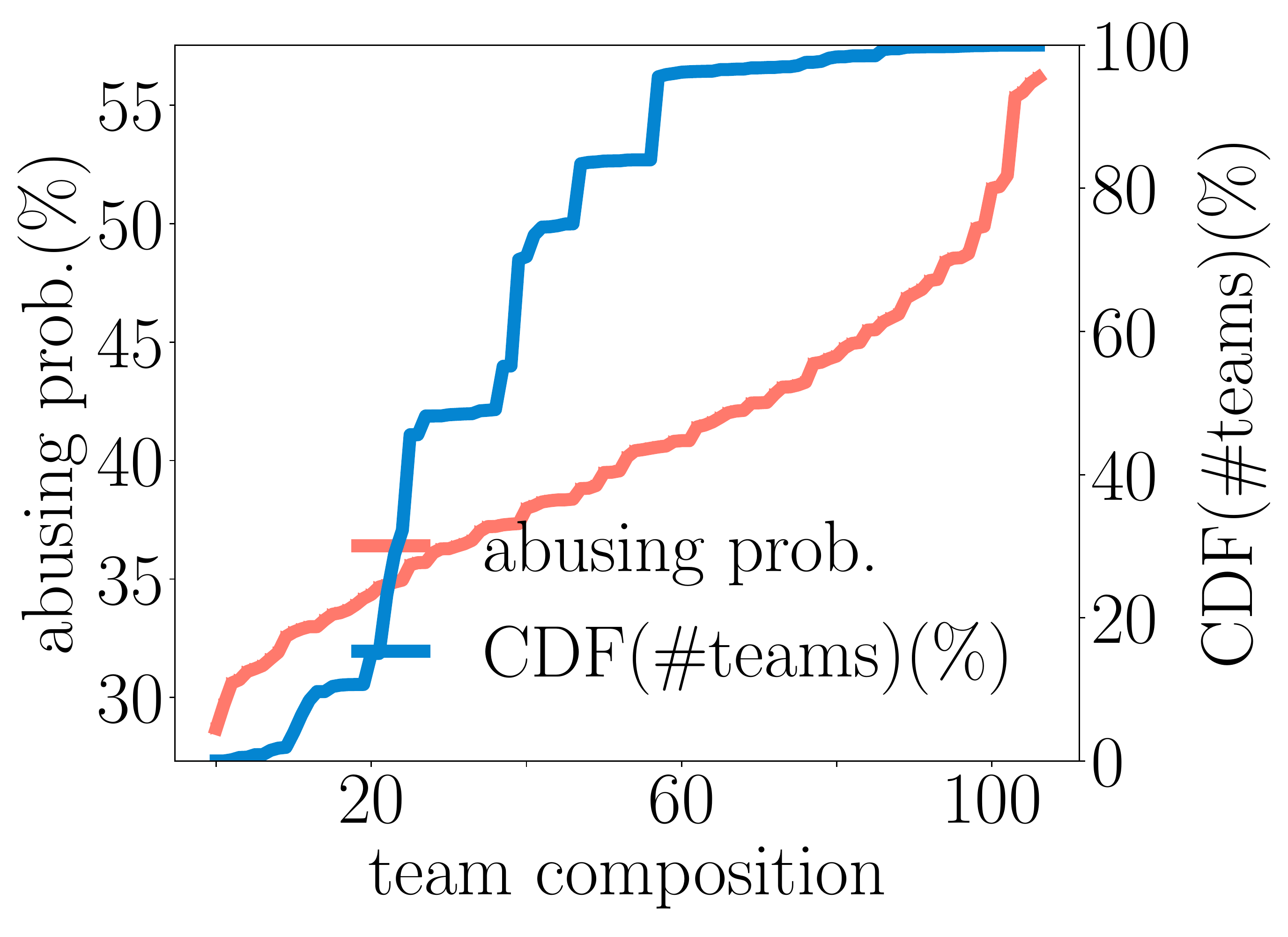}
		\caption{Sorted abusing probability.}
		\label{fig:abuse sorted}
	\end{subfigure}
	\begin{subfigure}[t]{0.32\textwidth}
		\centering
		\includegraphics[width=0.7\textwidth]{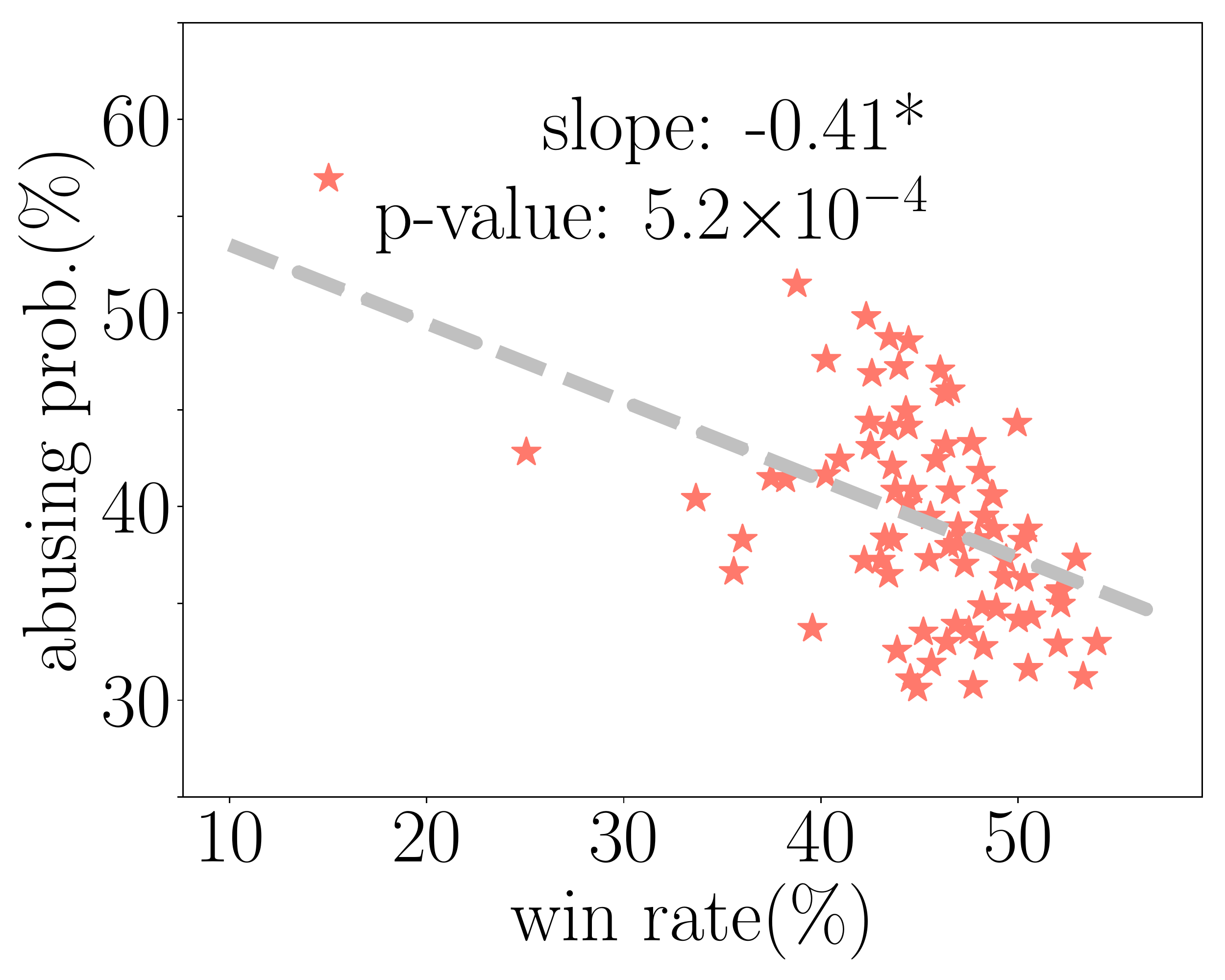}
		\caption{Abusing probability vs. winning rate.}
		\label{fig:winrate vs abuse}
	\end{subfigure}
	\begin{subfigure}[t]{0.32\textwidth}
		\centering
		\includegraphics[width=0.7\textwidth]{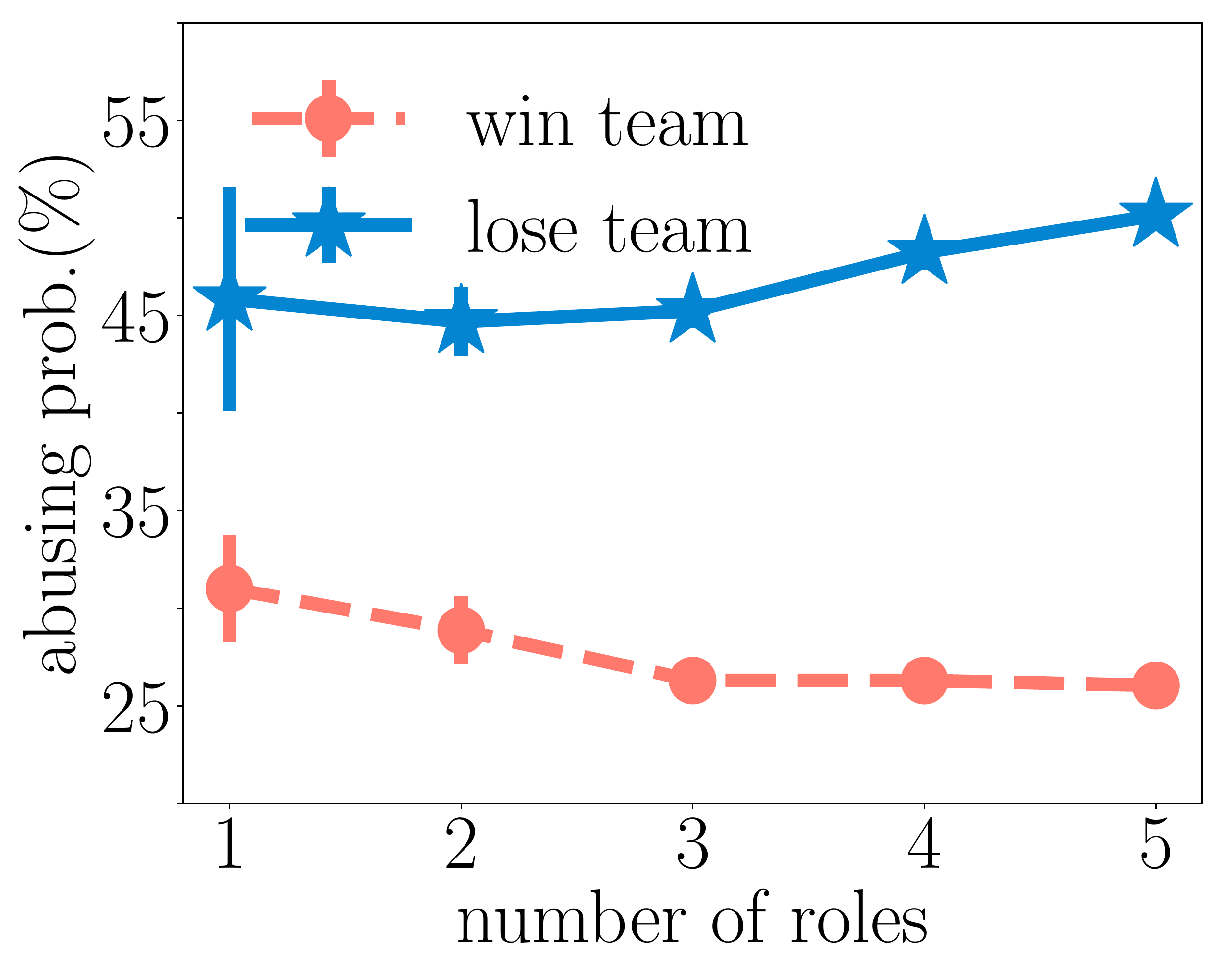}
		\caption{Abusing probability vs. team diversity.}
		\label{fig:diversity_abuse}
	\end{subfigure}
	\caption{Team composition and abusive language use. \figref{fig:abuse sorted} presents sorted abusing probability for different team compositions and its corresponding cumulative distribution function in \#teams. 
	In \figref{fig:winrate vs abuse}, each dot represents a team composition after filtering infrequent team compositions; $x$-value represents winning rate and $y$-value represents abusing probability. 
	In \figref{fig:diversity_abuse}, $x$-axis represents the number of roles in a team, and $y$-axis represents team-level abusing probability.
	} 
	\label{fig:abuse overview}
\end{figure*}

\section{Team Composition and Abusive Language Use}
\label{sec:abuse}
Our final measure of team effectiveness is concerned with the rapport in a team during the game.
We use abusive language use to capture the rapport.
Toxic behavior in online communities and gaming has received significant interests from our research community recently \cite{cheng2017,Cheng:ProceedingsOfIcwsm:2015,Kwak2015}.
Here we provide the first systematic study on the effect of team composition on team-level abusive language use.

\subsection{Team-level Abusive Language Use}
\label{sec:abuse:role}

\para{Team compositions vary in abusing probability (\figref{fig:abuse sorted}).}
In our dataset, we have a label of whether a message uses abusive language for all text messages based on a dictionary-based method officially used by Tencent.\footnote{Players may abuse using voice messages; here we only consider text messages.}
A team abuses if any player in that team abuses.
For each team composition, we define {\em abusing probability} as the fraction of games that this team composition abuses.

We find that similar to winning and surrendering, team compositions vary in abusing probability, ranging from 28.7\% to 56.2\%.
The cumulative distribution function of \#teams looks much more similar to surrendering in \figref{fig:surrender:sorted} than winning in \figref{fig:sorted_role_winrate}: most commonly used team compositions are neither the most nor the least abusive.
\tableref{tbl:roles_abuse} shows team compositions with the highest and lowest abusing probability. It is striking that all the top teams have multiple assassins ($\assassin$) and all the bottom teams have multiple supports ($\support$).
We will investigate further the interaction of individual-level abusing and team-level abusing in \secref{sec:abuse:individual}.

\para{Losing teams are more likely to abuse (\figref{fig:winrate vs abuse}).}
We hypothesize that losing teams are more likely to abuse because winning usually brings positive team morale, while losing leads to frustration and dissatisfaction \cite{Kwak2015}.
This is indeed the case as shown in \figref{fig:winrate vs abuse}.
Therefore, it is important to distinguish abusing probability between winning teams and losing teams.

\para{Diverse teams tend to abuse more when losing and abuse less when winning (\figref{fig:diversity_abuse}).}
We further explore the effect of role diversity on team-level abusing.
We observe different trends in winning teams and losing teams.
When a team wins, team diversity is associated with low abusing probability;
but it becomes the other way around if a team loses. 

\begin{figure*}[ht!]\setlength{\abovecaptionskip}{-0.05cm}\setlength{\belowcaptionskip}{-0.cm}
	\centering
	\begin{subfigure}[t]{0.254\textwidth}
		\includegraphics[width=\textwidth]{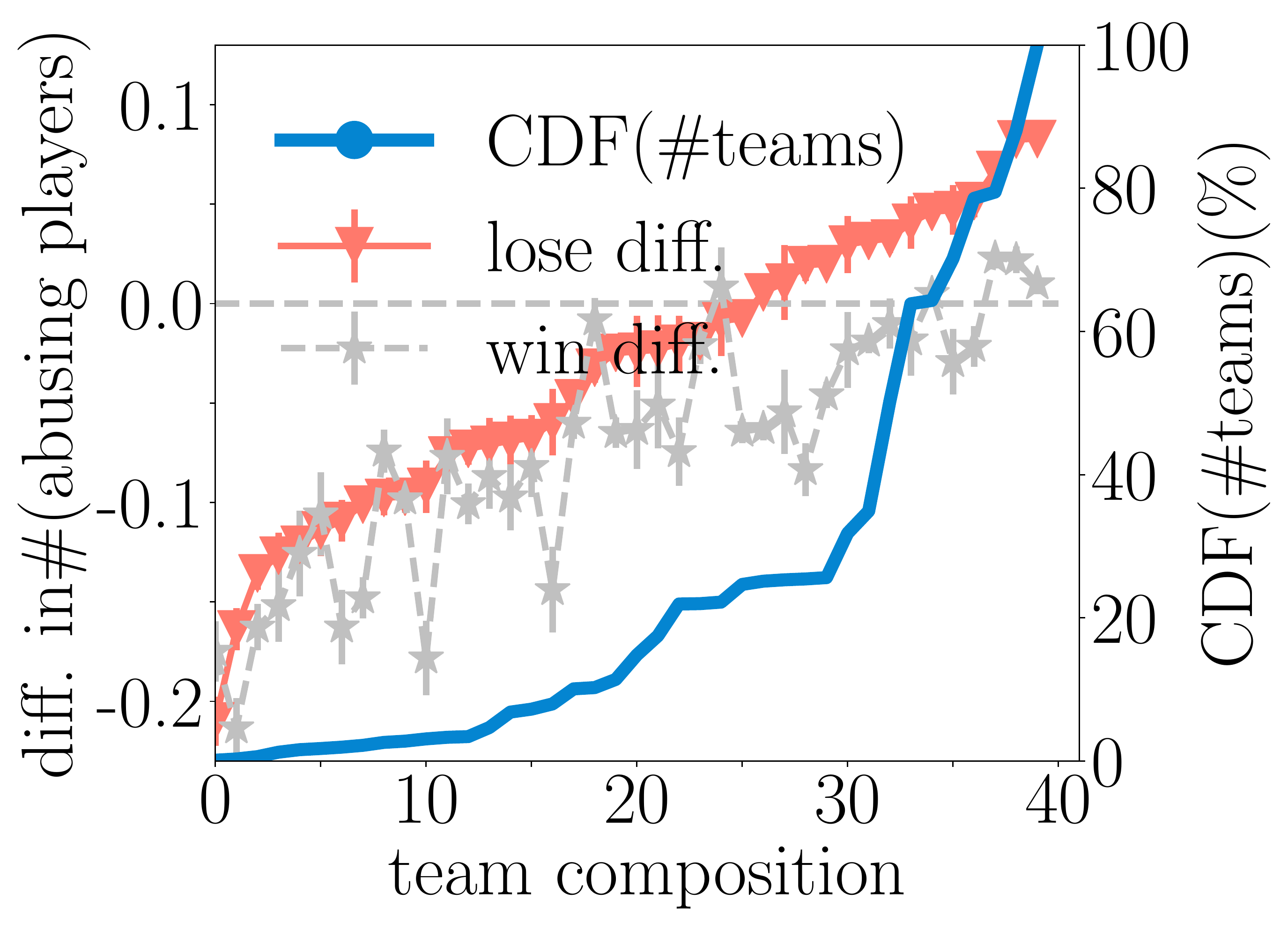}
		\caption{Observed \#abusing players $-$ expected values.}
		\label{fig:abuse_vs_expectation}
	\end{subfigure}
	\hfill
	\begin{subfigure}[t]{0.23\textwidth}
		\includegraphics[width=\textwidth]{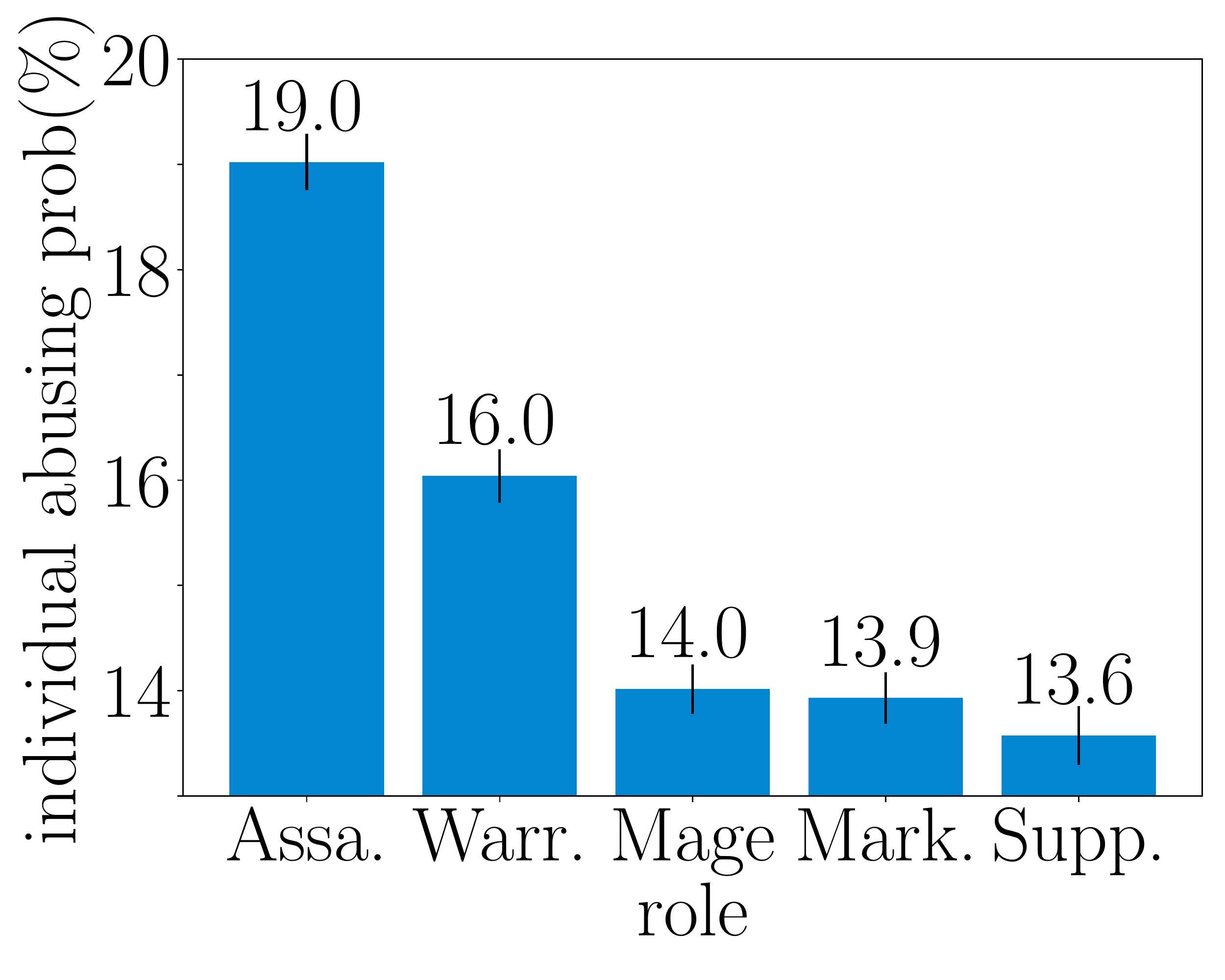}
		\caption{Abusing probability vs. roles.
		}
		\label{fig:abuse:role:prob}
	\end{subfigure}
	\hfill
	\begin{subfigure}[t]{0.23\textwidth}
		\includegraphics[width=\textwidth]{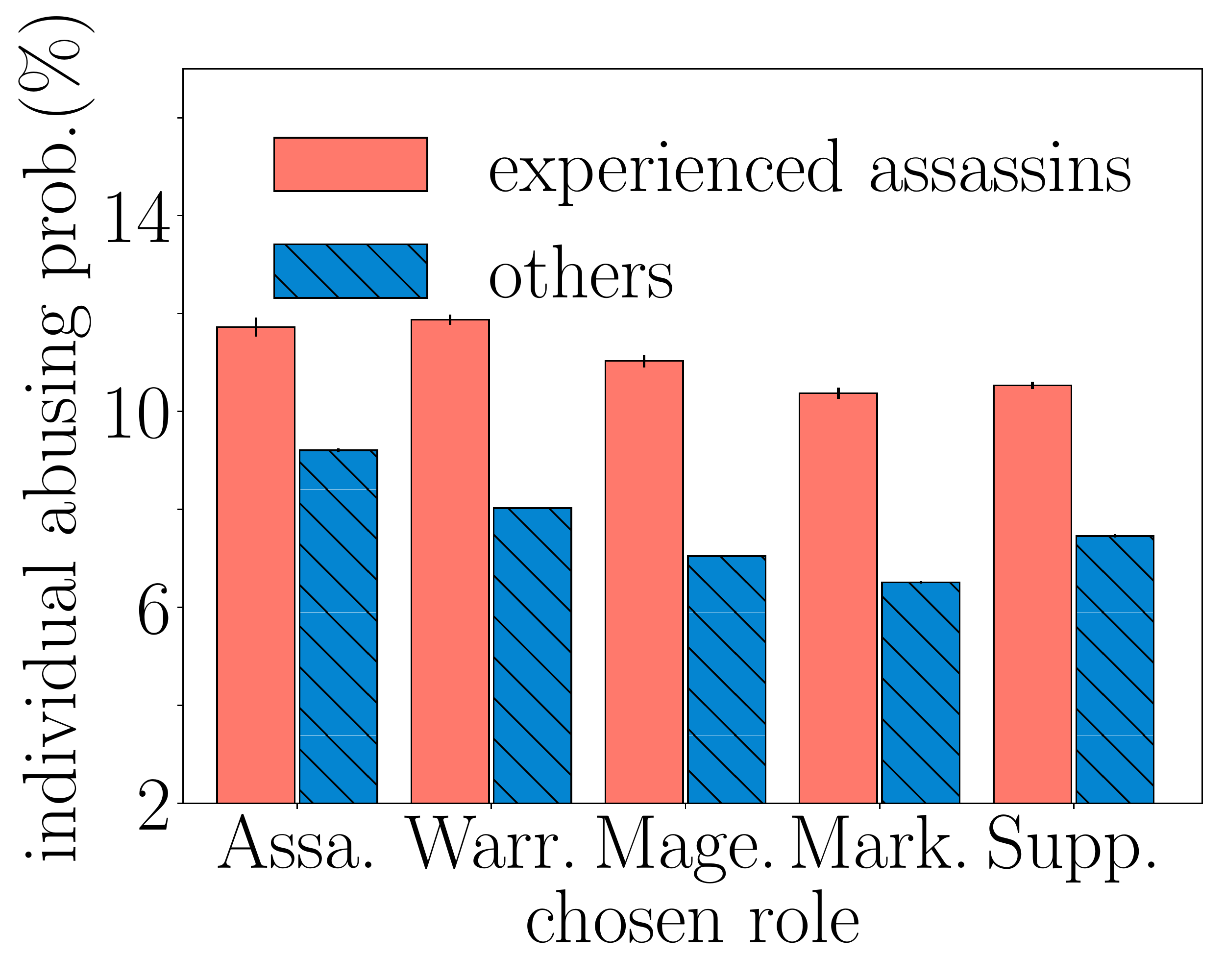}
		\caption{Abusing probability of the same roles by different players.}
		\label{fig:abuse:individual:assassin-abuse}
	\end{subfigure}
	\hfill
	\begin{subfigure}[t]{0.23\textwidth}
		\includegraphics[width=\textwidth]{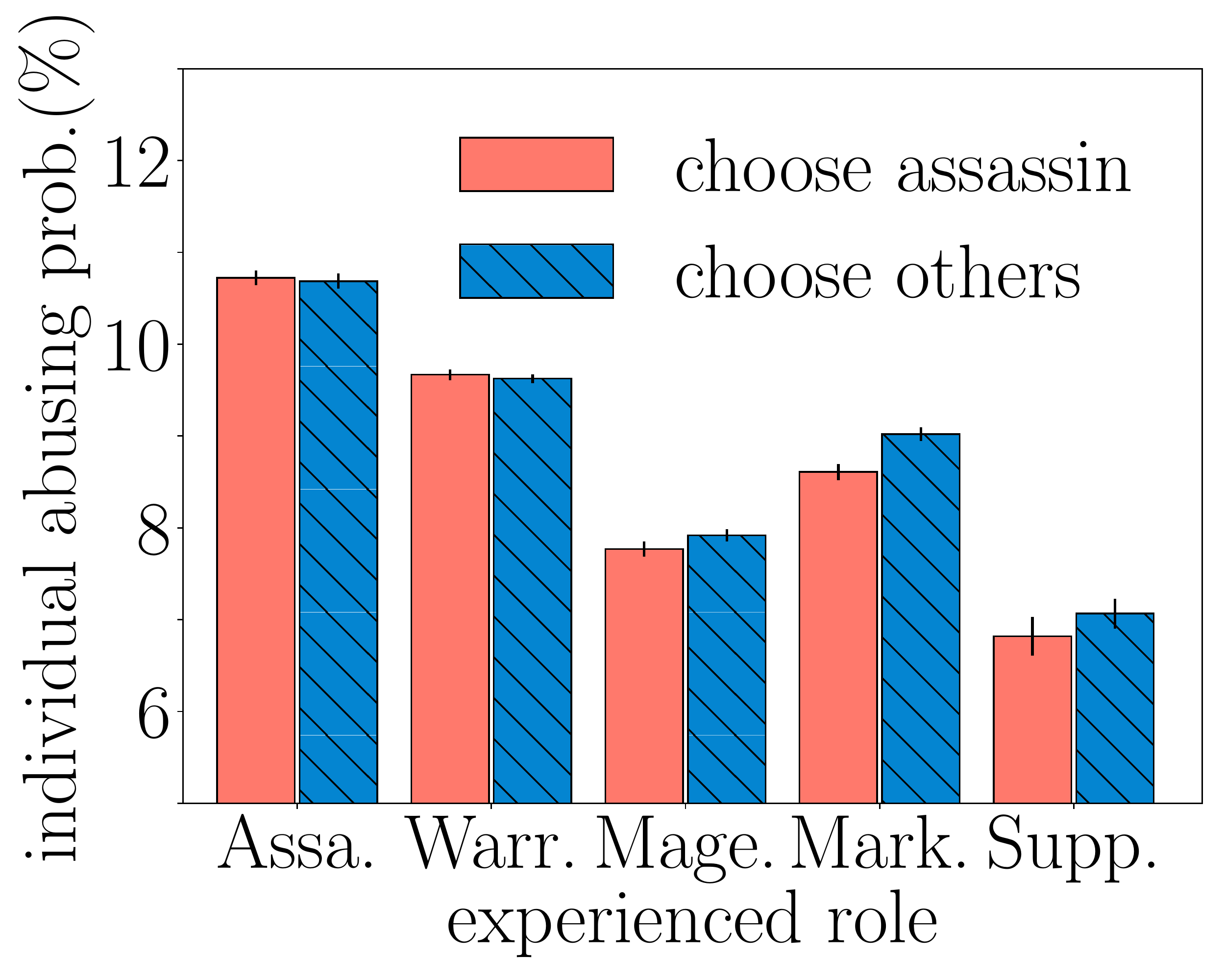}
		\caption{Abusing probability of the same player with different roles. 
		}
		\label{fig:abuse:individual:assassin-play}
	\end{subfigure}
	\caption{\figref{fig:abuse_vs_expectation} shows the difference between the observed number of abusing players and the expected value based on individual abusing probability.
		The right three figures investigate the ``abusive'' assassin phenomenon. 
		\figref{fig:abuse:role:prob} shows individual abusive probability grouped by roles.
		\figref{fig:abuse:individual:assassin-abuse} compares the abusing probability of players who usually play assassins with those who usually play other roles, grouped by their chosen role.
		\figref{fig:abuse:individual:assassin-play} compares the abusing probability when a player choose assassin with when the same player chooses other roles, grouped by his experienced role.
		Error bars represent standard errors.
	}
	\label{fig:abuse:role}
\end{figure*}

\subsection{Individual-level Abusive Language Use}
\label{sec:abuse:individual}

Different from winning and surrendering, abusive language use is an individual behavior.
It provides a great opportunity to understand the effect of team composition on individual behavior and shed light on the ``situation vs. personality'' debate  \cite{person-situation-09,Kenrick:AmericanPsychologist:1988}.
Note that we conduct this part of experiments on players who have played at least 20 games to ensure sufficient samples for statistics.  

\para{A team does not equal the sum of individuals (\figref{fig:abuse_vs_expectation}).}
We define individual abusing probability based on the fraction of games that a player abuses in.\footnote{All individual abusing probability is based on individual level samples. Individual samples can be grouped based on criteria other than all games played by a single player, e.g., we compute individual abusing probability for all winning (losing) games of a player in \figref{fig:abuse_vs_expectation} and for all players who choose a particular role in \figref{fig:abuse:role:prob}.
}
Because of linearity of expectation, the total number of expected abusing players in a team is simply the sum of each player's abusing probability:
\begin{equation}
\label{exp of abuse scale}
\mathbb{E}(\text{\#abusing players in a team}) = 
\sum_{v \in \mathtt{team}}P_{\mathtt{abuse}}(v)
\end{equation}

\noindent where $P_{\mathtt{abuse}}$ is estimated separately when a player wins or loses because abusing is associated with losing.
This expectation is only valid if all players act independently in a team, but studies on teamwork have shown that a team does not equal the sum of individuals \cite{belbin2012team}. 
Therefore, we examine the discrepancy between observed values and expected values to understand how team composition influences individual players. 
This analysis is done on prediction games since it requires historical information of every team member.

Figure~\ref{fig:abuse_vs_expectation} shows the difference between observed and expected values for winning and losing teams,
where \textit{win/lose diff.} refers to the disparity between observed and expected number of abusing players($\mathbb{O}(\text{\#abusing players}) - \mathbb{E}(\text{\#abusing players})$).
For most team compositions, the difference between observed values and expected values is not zero, indicating that individuals abuse differently depending on team compositions.
In particular, in the most commonly used teams (when the CDF grows quickly on the right of the plot), individuals are more likely to abuse than expected when losing. 

\para{Players who prefer leading roles are more abusive (\figref{fig:abuse:role:prob}, \ref{fig:abuse:individual:assassin-abuse}, \ref{fig:abuse:individual:assassin-play}).} 
There always exists a leader or a major contributor in a team.
In \kingofglory,
assassins usually take this role in a team because of their explosiveness: assassins can carry the team and control game pace.
A failed assassin may lead the team to lose.
 
\figref{fig:abuse:role:prob} shows that assassins are more likely to abuse than other roles.
We call this the ``abusive assassin'' phenomenon.
One natural question arises: are assassins more abusive because abusive players tend to choose assassins, or players become more abusive when choosing assassins?

To answer this question, we compare the abusing probability of the same roles chosen by different players and the same player choosing different roles.
We define the {\em experienced role} for a player 
if that player chooses a role frequently, i.e., playing in more than $50$\% of games. 
This procedure identifies experienced assassin players, experienced warrior players, etc.
\figref{fig:abuse:individual:assassin-abuse} presents the abusing probability of experienced assassin players when they choose each role (the same player choosing different roles).
We observe that these experienced assassin players are much more likely to abuse than other players no matter what role they choose.

\figref{fig:abuse:individual:assassin-play} examines the alternative hypothesis: players become more abusive when choosing assassins.
We compare the same player's abusing probability when they choose assassins with choosing other roles, grouped by their experienced roles.
We find that players do not become more abusive when choosing assassins.
If anything, experienced mages and experienced marksmen are actually more abusive when they choose roles other than assassins.

Overall, our results in \kingofglory support the hypothesis that abusive players tend to choose assassins, the leading role in a team, and players do not become abusive when choosing assassins.

\begin{figure}[t]\setlength{\abovecaptionskip}{-0.cm}\setlength{\belowcaptionskip}{-0.cm}
	\centering
	\begin{subfigure}[t]{0.225\textwidth}
		\includegraphics[width=\textwidth]{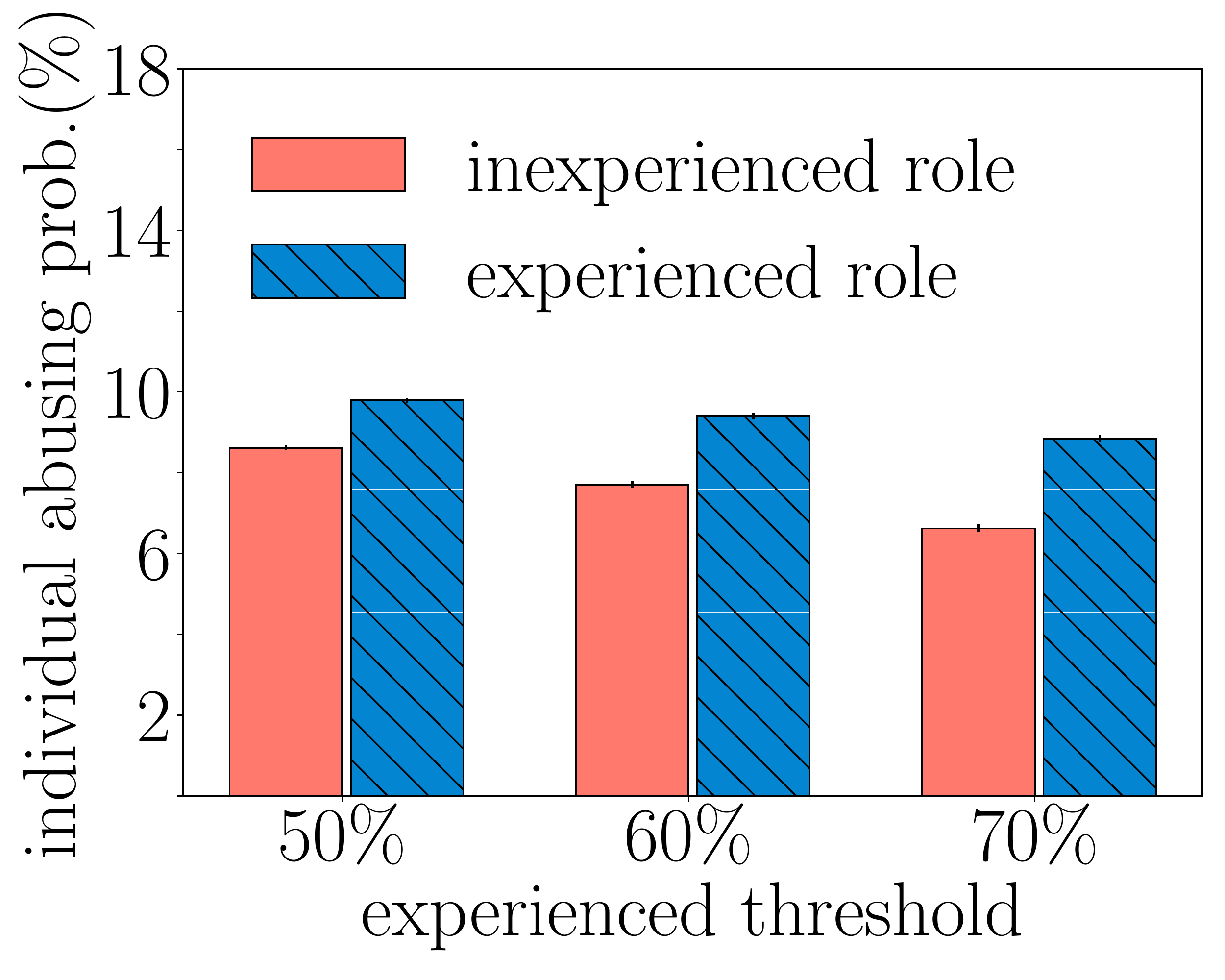}
		\caption{Abusing probability vs. (in)experienced roles.}
		\label{fig:abuse:familiar:player}
	\end{subfigure}
	\hspace{0.01\textwidth}
	\begin{subfigure}[t]{0.225\textwidth}
		\includegraphics[width=\textwidth]{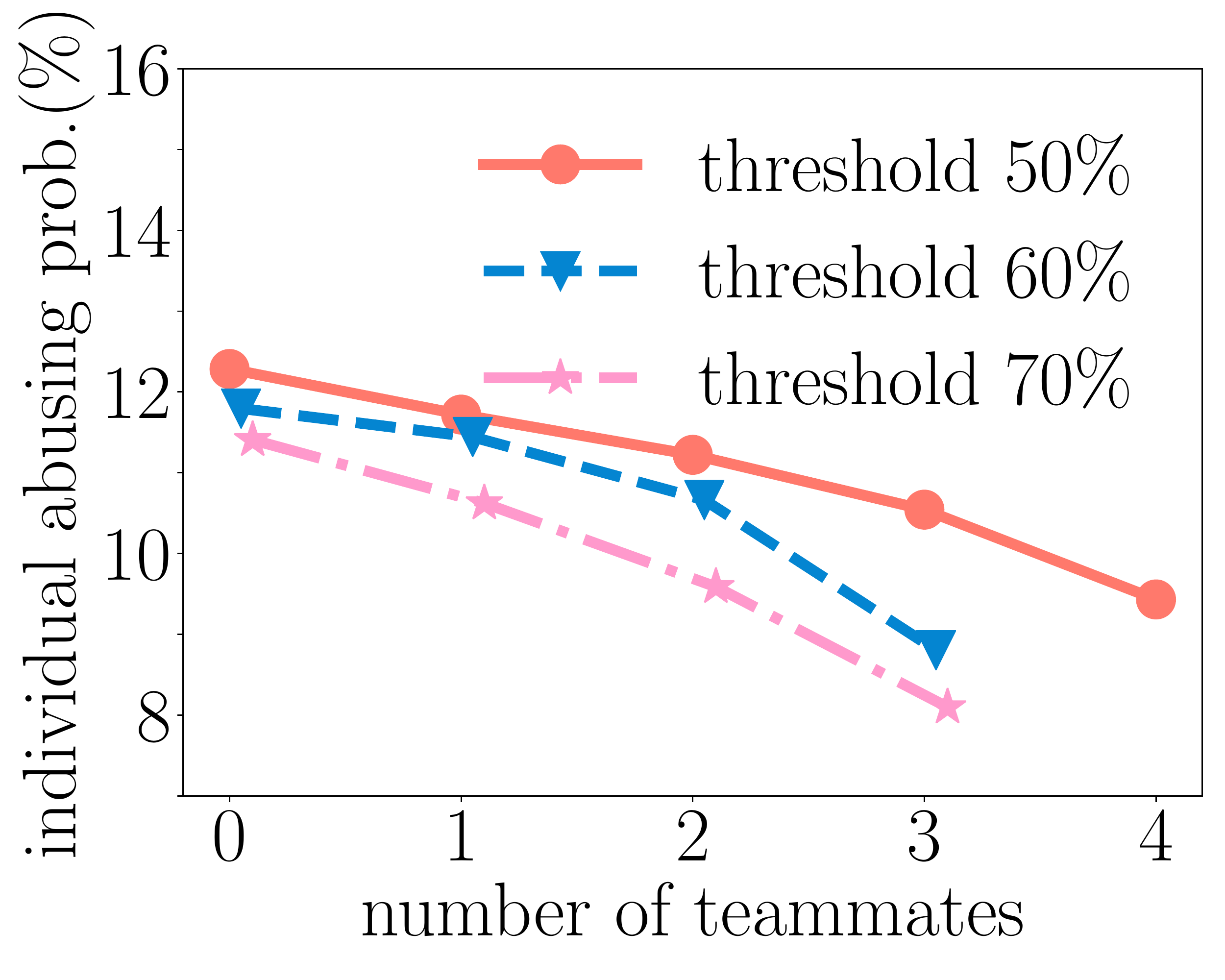}
		\caption{Abusing probability vs. \#experienced roles in teammates.
		}
		\label{fig:abuse:familiar:team}
	\end{subfigure}
	\caption{Individual abusing probability and experienced roles.
	Fig.~\ref{fig:abuse:familiar:player} compares the abusing probability when a player plays his experienced role with when playing inexperienced roles.
	In Fig.~\ref{fig:abuse:familiar:team},
	$x$-axis denotes the number of experienced roles (defined by given threshold) among teammates, while $y$-axis denotes individual abusing probability.
	}
	\label{fig:abuse:experience}
\end{figure}

\para{Experienced players are more likely to abuse, but not when they play with other experienced players (\figref{fig:abuse:experience}).}
Studies have shown that experienced individuals tend to abuse novices \cite{honeycutt2005hazing}. 
We hypothesize that a player is more likely to abuse when playing his experienced role.
In addition to using 50\% as a threshold to define experienced roles, we also use 70\% and 60\% here.
\figref{fig:abuse:familiar:player} shows that players are more abusive when playing their experienced roles no matter what threshold is.

We next further explore how players are influenced by teammates when choosing experienced roles. 
\figref{fig:abuse:familiar:team} shows that as the number of experienced roles increases among a player's teammates, his individual abusing probability player
decreases (since this analysis requires historical information of every team member, it is done on prediction games).
This observation indicates that a player becomes more likely to abuse when choosing experienced roles partly because other less experienced players do not meet this experienced player's expectation and cause frustration.

%% file: exp.tex
\begin{figure*}[t]
	\centering
	\begin{subfigure}[t]{0.23\textwidth}
		\centering
		\includegraphics[width=\textwidth]{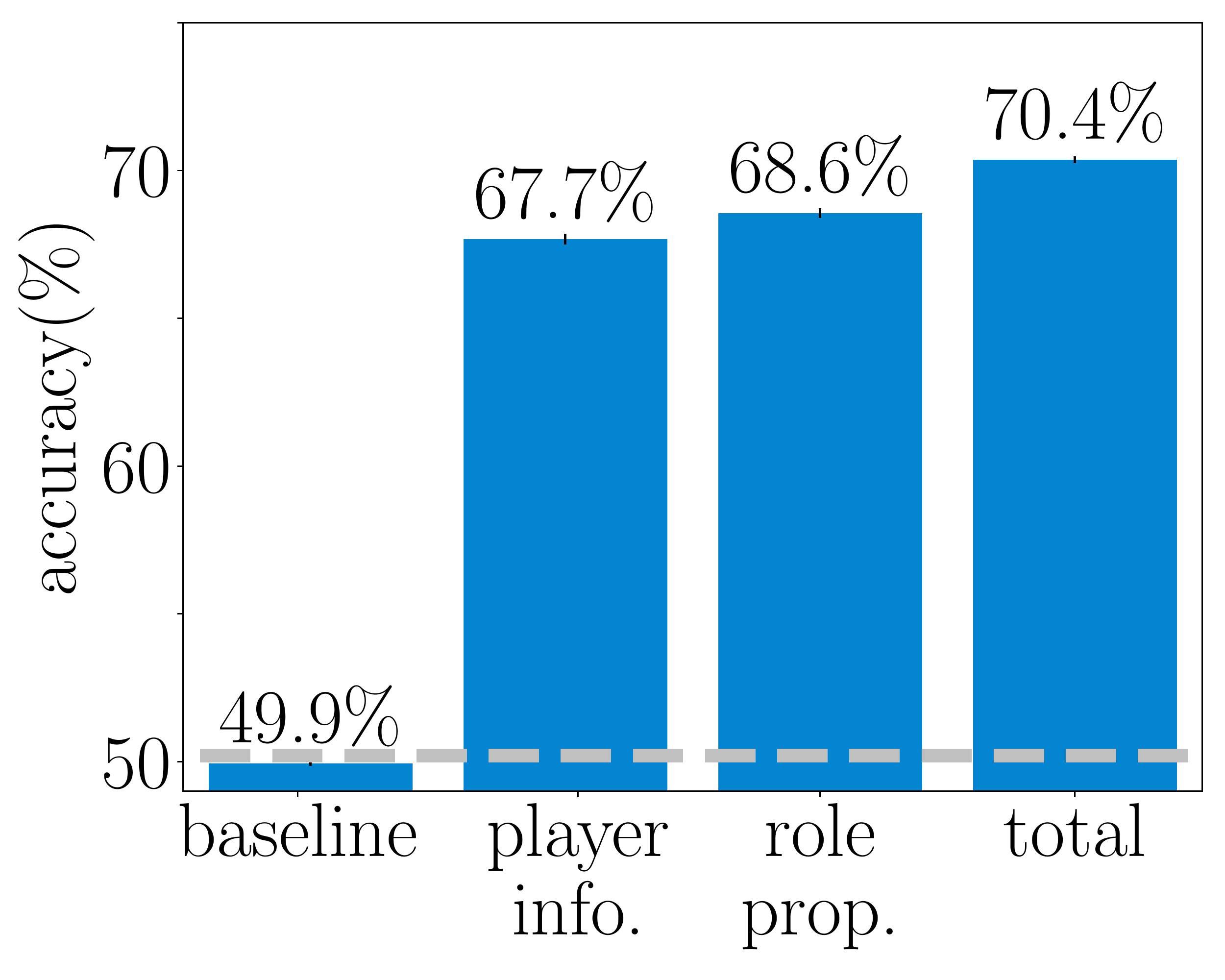}
		\caption{Winning team prediction}
		\label{fig:predict:winlose-rank}
	\end{subfigure}
	\hfill
	\begin{subfigure}[t]{0.23\textwidth}
		\centering
		\includegraphics[width=\textwidth]{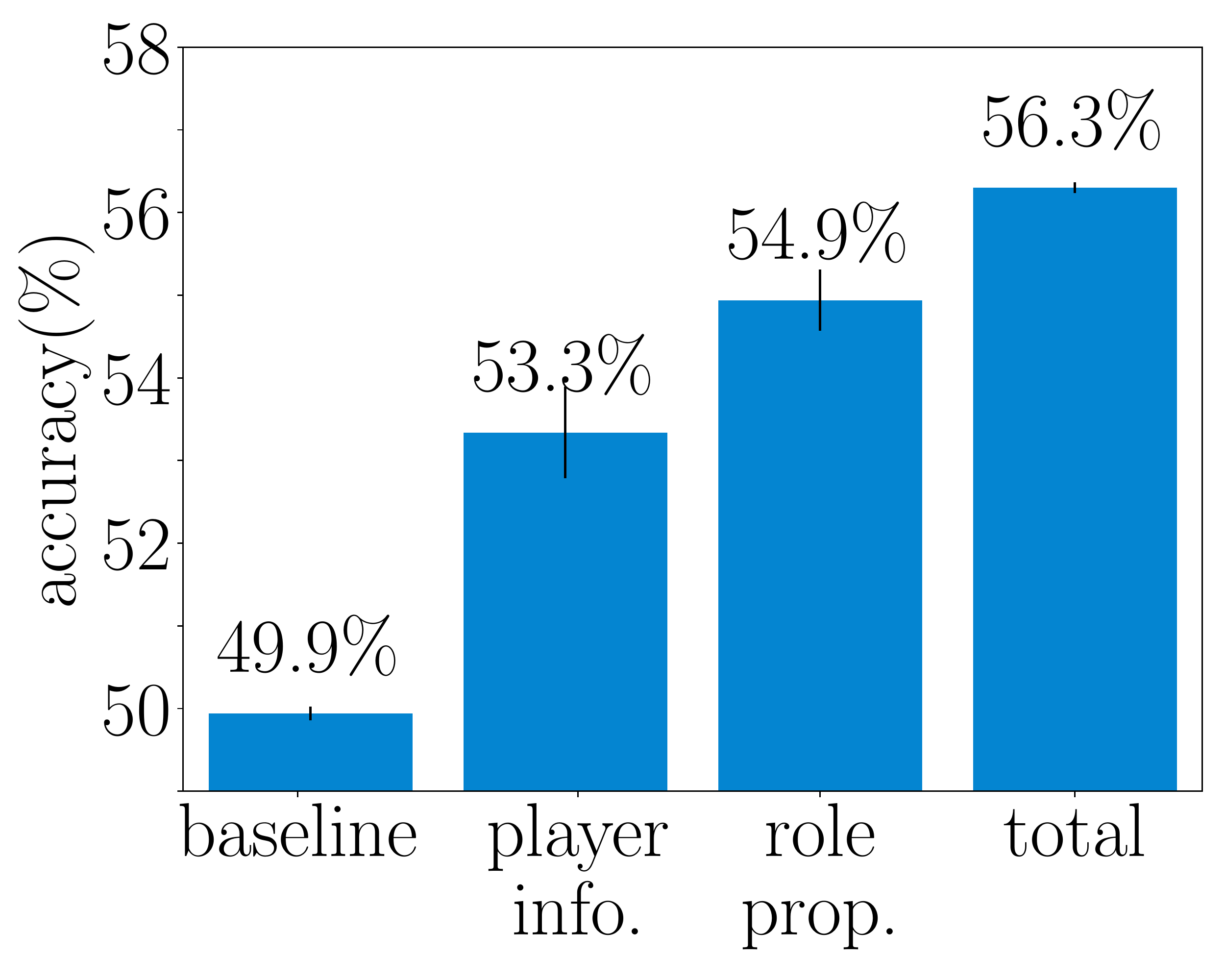}
		\caption{Winning team prediction on games where the rank gap $\delta=0$}
		\label{fig:predict:winlose-norank}
	\end{subfigure}
	\hfill
	\begin{subfigure}[t]{0.23\textwidth}
		\centering
		\includegraphics[width=\textwidth]{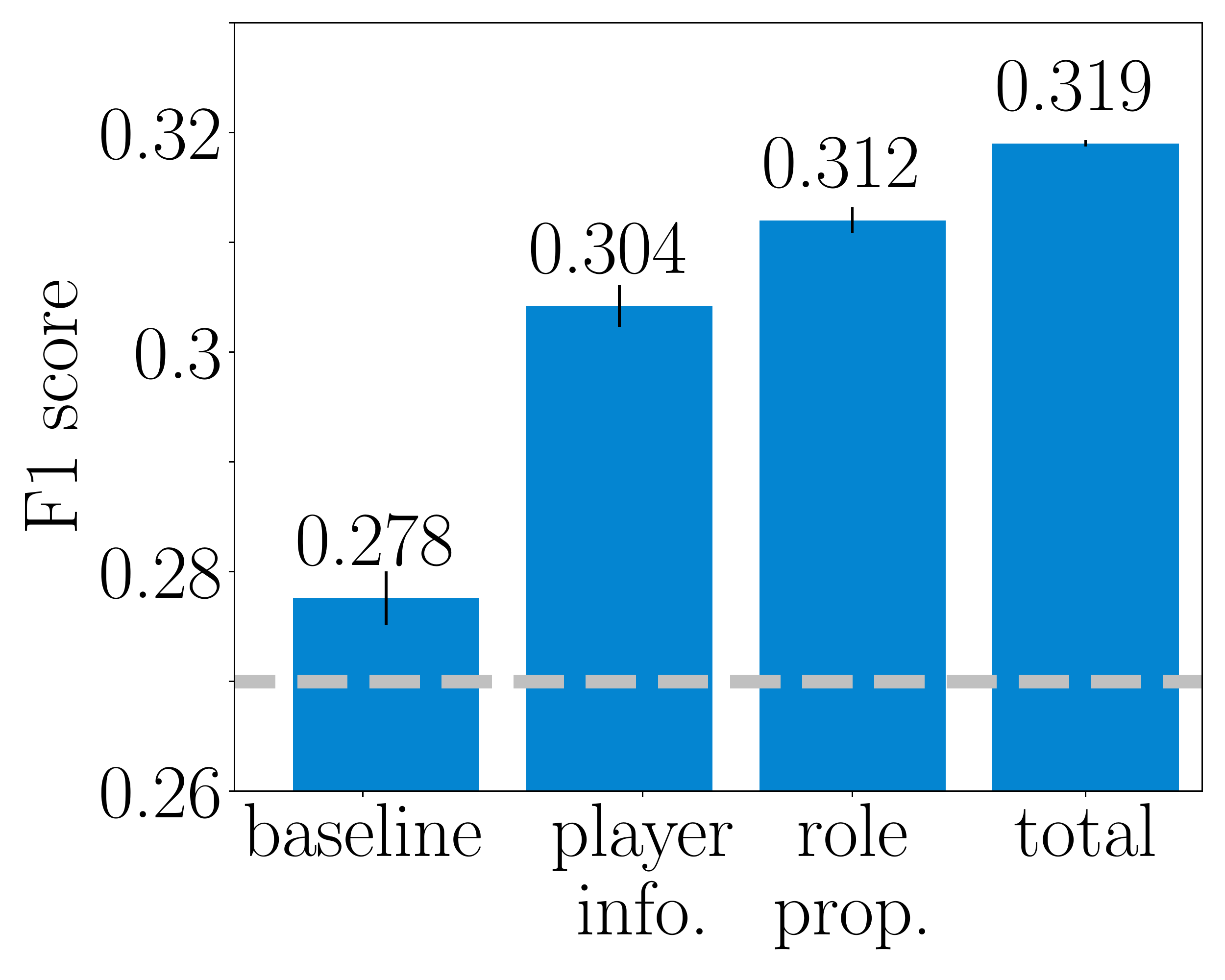}
		\caption{Surrender prediction}
		\label{fig:predict:surrender}
	\end{subfigure}
	\hfill
	\begin{subfigure}[t]{0.23\textwidth}
		\centering
		\includegraphics[width=\textwidth]{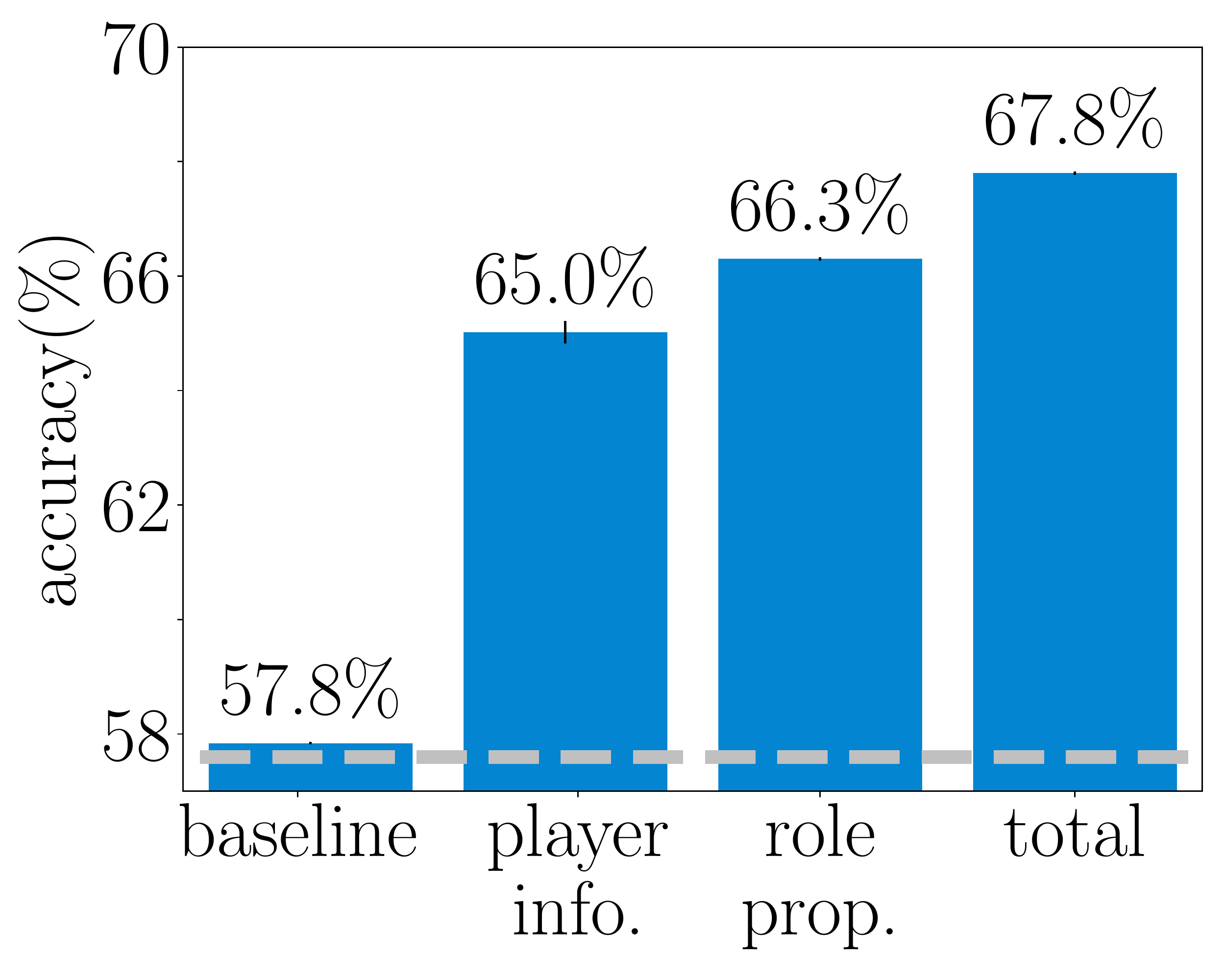}
		\caption{Abuse prediction}
		\label{fig:predict:abuse}
	\end{subfigure}

	\caption{Prediction results. Gray lines indicate the performance of a majority baseline in \figref{fig:predict:winlose-rank} and \ref{fig:predict:abuse}, and a random baseline in \ref{fig:predict:surrender}. Features that incorporate role combinations with player information (role prop.) consistently outperform others. 
	}
	\label{fig:predict:all}
\end{figure*}

\section{Prediction Experiments}
\label{sec:exp}

In this section, we set up three prediction tasks for winning, surrendering, and abusing respectively to further examine the effect of team composition in a predictive setting.
Overall, we find that features based on team composition consistently outperform features based on individual players.

\subsection{Experiment Setup}
\label{sec:exp:setup}

\para{Prediction tasks and evaluation metrics.} 
Our goal is to evaluate whether modeling team composition brings extra predictive power to modeling individual players.
Because team composition cannot be altered after a game starts, we focus on making predictions before a game starts and consider the following three prediction tasks.
Our results can potentially inform both players in choosing roles and the matching-making system to match players.

\begin{itemize}[leftmargin=*,noitemsep,topsep=0pt]
\item Winning team prediction\cite{xia2017contributes}.
This task aims to predict which team will win before the game starts.
We randomly swap the order of two teams and label a game positive if the first team wins and negative if the second team wins.
This leads to a balanced dataset, so we use accuracy for evaluation.

\item Surrender prediction.
This task aims to predict whether the losing team surrenders in a game.
Only 8K games (10.5\%) end with surrender, so we use F1 as evaluation metric in this task.

\item Abuse prediction.
Before a game starts, this task predicts whether abusive language use happens in the game.
Abusive language use happens in 44K games (57.93\%), so we use accuracy for evaluation in this task.
\end{itemize}

\para{Feature sets.} We consider the following feature sets.

\begin{itemize}[leftmargin=*,noitemsep,topsep=0pt]
\item Baseline.
People's mood varies over time and is associated with abusive language use \cite{cheng2017}. Our baseline features include day of the week and hour of the day encoded as one-hot vectors.

\item Player information.
Depending on the task, we compute each player's rank level,  
winning rates in the last 10 ranked games, surrender probability, and abusing probability from \textit{historical games}.
We use the average, max, min, and standard deviation of those properties of all players in a team as features.

\item Role properties. 
A role combination of $k$ refers to all possible combinations of $k$ roles. For instance, one-role combinations refer to the five roles in \tableref{tb:role_character}, two-role combinations refer to 15 possible combinations of two roles ($\binom{5}{1} + \binom{5}{2}$).
We further incorporate player information in role combinations of both teams.
For each role combination in a team, we compute the average of \textit{role-based} player information 
in an instance of that combination and then use the average, max, min, and std of all instances in that combination as features. 
Taking an example to further explain role-based player information, 
the feature \textit{winning rate of role $c$ for player $v$} is defined as the probability of $v$ wins when she plays $c$.
We only consider up to two-role combinations ($k \leq 2$) when comparing prediction performances.
Later we will explore the effect of considering three-role combinations up to five-role combinations, i.e., full team compositions.
\end{itemize}

We also use the union of all feature sets (``total'').
Because we find that game duration plays an important role in surrender, we add game duration to every feature set for surrender prediction.
In abuse prediction, we do not distinguish two teams.

\para{Prediction setup.}
For each task, we conduct 5-fold nested cross-validations on \textit{prediction games} in the third week and player information is extracted from \textit{historical games} in the first two weeks. 
We use \textit{Xgboost}~\cite{chen2016xgboost} since it is often the winning solution in Kaggle competitions and shows strong performance compared to other methods such as logistic regression in our preliminary experiments. 
We grid search hyperparameters in max depth (\{1, 3, 5, 7, 9\}), learning rate (\{0.1, 0.2\}), and number of estimators (\{100, 200, 300\}).
\subsection{Experiment Results}
\label{sec:exp:res}

\para{Prediction performance (\figref{fig:predict:winlose-rank}, \ref{fig:predict:winlose-norank}, \ref{fig:predict:surrender}, and \ref{fig:predict:abuse}).}
In all prediction tasks, we observe a consistent pattern that $\textit{total} > \textit{role property}$
$> \textit{player information} 
> \textit{baseline}$.
Player information achieves relatively better performance than baselines since individual characteristics definitely influence team effectiveness in some degree.
By incorporating player information with role combinations, role properties consistently outperform player information.
The extra predictive power from modeling team composition can be reflected by the difference between \textit{total} and player information: 
e.g., \textbf{70.4\%} vs 67.7\% in winning prediction for all games, and \textbf{67.8\%} vs 65.0\% in abusing prediction. 
In winning prediction, the effect of team composition is further amplified when the rank gap being controlled:  
\textbf{56.3\%} vs 53.3\%.  
The difference is the least prominent in the prediction of surrender, as this behavior is often highly influenced by teammates and various situations during games,
while role properties still improve the performance by \textbf{1.5\%} in surrender prediction.  

Recall that \figref{fig:rankgap distribution} shows a clear influence of rank level on winning rate. 
Therefore, to further understand the effect of team compositions, 
we report the performance on games where the rank level gap $\delta=0$ in \figref{fig:predict:winlose-norank}.
We find that after the rank level gap is controlled, this task becomes much more challenging and the performance of all feature sets drop significantly compared to \figref{fig:predict:winlose-rank}.
However, the difference between role properties and player information almost doubles (1.6\% vs. 0.9\%), demonstrating the effectiveness of incorporating team compositions, especially when the two teams are similarly skilled.
In contrast, rank level does not affect the performance when predicting surrender or abuse.

\begin{figure}[tp!]
	\centering
	\begin{subfigure}[t]{0.23\textwidth}
		\includegraphics[width=\textwidth]{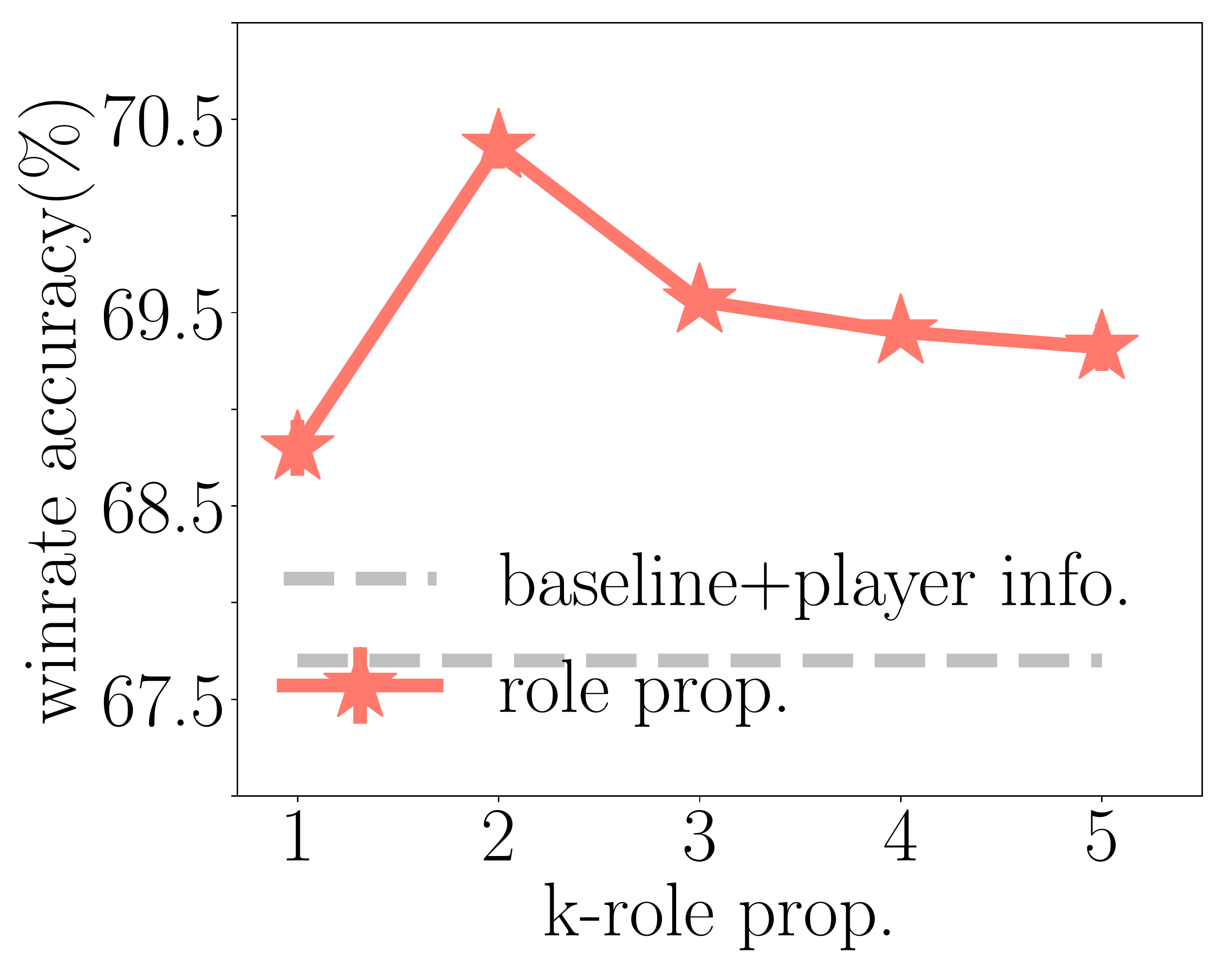}
		\caption{Winning team prediction.}
		\label{fig:n-gram-winrate}
	\end{subfigure}
	\hfill
	\begin{subfigure}[t]{0.23\textwidth}
		\includegraphics[width=\textwidth]{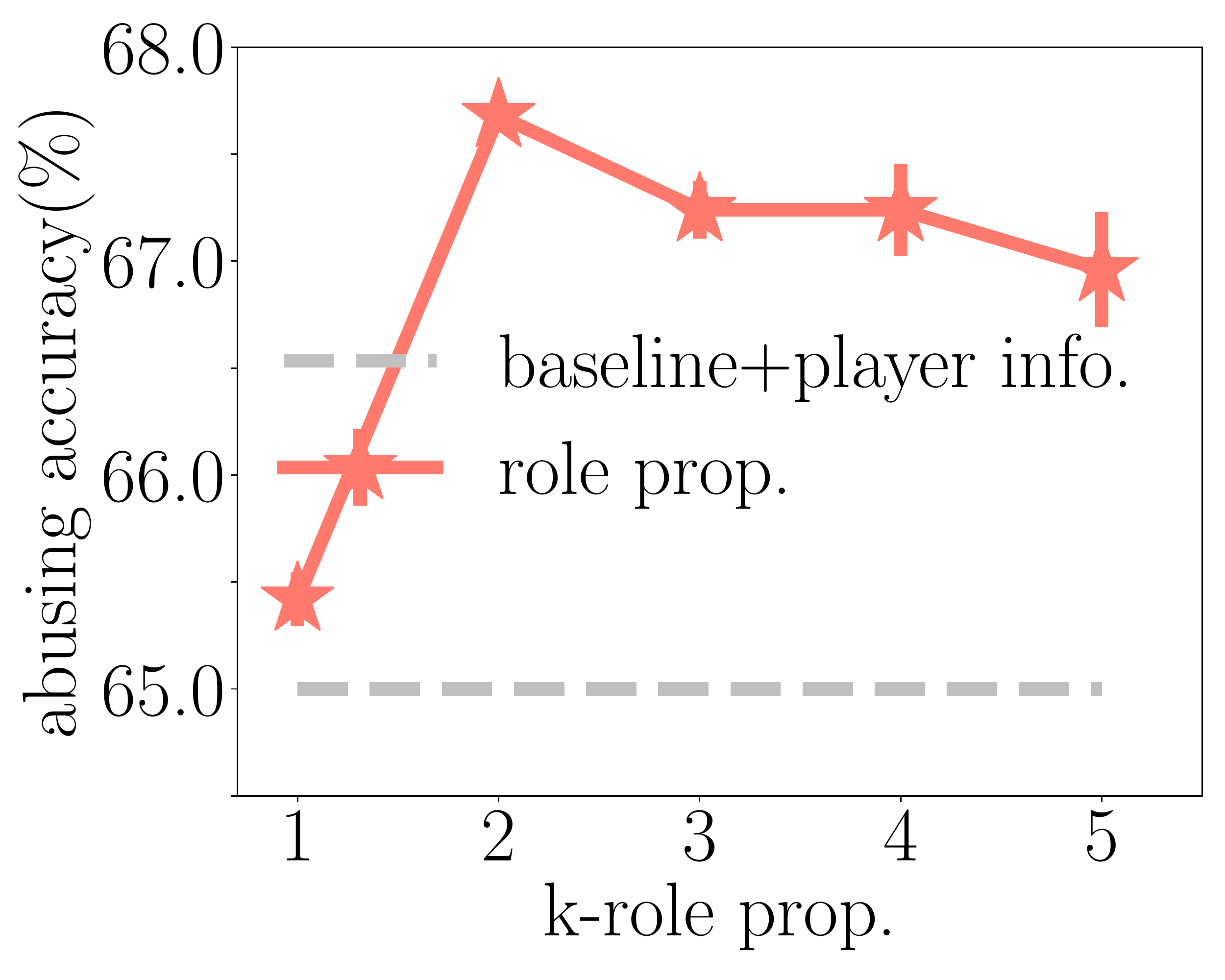}
		\caption{Abuse prediction.}
		\label{fig:n-gram-abuse}
	\end{subfigure}
	\caption{Effect of role combinations. It shows that modeling two-role combinations is sufficient to capture the effect of team composition.}
	\label{fig:n-gram}
\end{figure}

\para{Modeling two-role combinations is sufficient to capture the effect of team composition (\figref{fig:n-gram}).}
To further illustrate the effect of team compositions, we examine the performance of adding $k$-role combinations to player information and baseline one by one:
we first add single-role features, then add features based on two-role combinations, until adding features based on entire team compositions.
Taking winning team (on all games) and abuse prediction as examples, 
\figref{fig:n-gram} shows that the performance improvement saturates after adding two-role combinations. 
It suggests that modeling single roles and two-role combinations provides most of the predictive power.

\begin{table}[t]\setlength{\abovecaptionskip}{-0.cm}\setlength{\belowcaptionskip}{-0.cm}
	\centering
	\small
	\begin{tabular}{l | l | r}
		\toprule
		rank & features & importance(\%)\\
		\hline
		$\#7$ & minimum winrate of $\mage-\mage$ & $1.1$\\
		$\#8$ & minimum winrate of $\assassin-\marksman$ & $1.1$\\
		$\#9$ & minimum winrate of $\assassin-\warrior$ &$1.0$\\
		$\#10$ & std(rank level) of $\mage$ & $1.0$\\
		$\#11$ & average rank level of $\marksman$ & $0.9$\\
		\toprule
	\end{tabular}
	\caption{Top five role property features in winning team prediction. 
		Feature importance is defined as the frequency of a feature being used in a tree.}
	\label{tb:winning feature}
	\vspace{-0.9cm}
\end{table}

\para{Feature analysis.}
We finally explore the importance of features 
by using winning team prediction task as an example. 
When combining role properties and player information, 
average rank levels of two teams are the most important features (ranked top-2), as
well as average individual winning probability (in top-5).
To understand the key features in role properties,
we show the top five features in role properties in \tableref{tb:winning feature}.
Consistent with the characteristics of roles 
where mage ($\mage$), marksman ($\marksman$) and warrior ($\warrior$) are most commonly used, 
the combinations of these roles are more important in role properties.
And as marksman ($\marksman$) and mage ($\mage$) are the key roles to cause damages, 
role combinations related with these roles are particularly significant. 
The first three of the top five are all two-role features, which again
indicates that modeling two-role combinations is effective to capture the team performance.

%% file: relate.tex

\section{Related Work}
\label{sec:related}
In addition to studies mentioned throughout the paper, we discuss additional related work in three strands.

\para{Team formation.}
Researchers from the data mining community have formulated the problem of team formulation as a constrained optimization problem based on each individual's skills and their social networks \cite{Anagnostopoulos:ProceedingsOfCikm:2010,anagnostopoulos2012online,Li:IeeeTransKnowlDataEng:2017,Lappas:ProceedingsOfKdd:2009}.
For instance, \citet{anagnostopoulos2012online} take into account of the task requirements, individual workloads, and team communication costs.
\citet{Li:IeeeTransKnowlDataEng:2017} study a family of problems in team enhancement including team member replacement, team expansion, and team shrinkage.

\para{Social/team roles.} Digitalization of human traces have increasingly made implicit social/team roles explicit and enabled large-scale study on theories of social/team roles \cite{meredith2011management,belbin2012team,biddle1986recent,Yang:ProceedingsOfIcwsm:2016}. 
\citet{yang2015rain} study how social roles influence the process of online information diffusion. 
\citet{Danescu-Niculescu-Mizil:2012:EPL:2187836.2187931} demonstrate changes in language use after a Wikipedia user becomes an administrator.
\citet{Yang:ProceedingsOfIcwsm:2016} infer editor roles on Wikipedia based on behavior traces. 
In the context of gaming, \citet{birgit2011} examined problematic gaming behavior and depressive tendencies among people who play different types of online-games. 

\para{Online gaming.}
Many researchers have recognized that online gaming, 
especially MOBA and MMOGs (Massively Multiplayer Online Games),
can serve as a platform for studying individual and team behavior, group norms, and communities \cite{Ducheneaut:ProceedingsOfChi:2007,Kim:CommunicationsOfTheAcm:}.
In such games, players not only develop individual skills, 
but also coordinate and communicate with others. 
The existence, significance, and influence of such social interactions have been examined in many studies~ 
\cite{ducheneaut2006alone,hudson2014measuring,
	losup2014analyzing,freeman2018multiplayer,szell2010measuring}. 

Team performance is the most heavily studied topic among our three measures of effectiveness.
\citet{pobiedina2013ranking,pobiedina2013successful} explore factors that influence player's performance in Dota 2
and reveal that team-level interactions, especially the number of friends in a team, relate to team performance. 
Similar to our work, prior studies such as \cite{agarwala2014learning,conley2013does} 
demonstrate that different role combinations
may greatly influence team performance. 
Deep learning based recommendation systems are also built to recommend heroes or roles to players\cite{sapienza2018deep}. 

Furthermore, gamification can also potentially play an important role in education and scientific discovery \cite{khatib2011algorithm}.
In particular, \citet{petter2017} suggests that researchers should embrace the growing popularity of online gaming and seek opportunities in studying online gaming in the context of education, individual, and team behavior.
In addition to these exciting opportunities, online gaming leads to issues such as addiction and depression \cite{birgit2011,kuss2012online}.

%% file: conclude.tex
\section{Conclusion}
\label{sec:conclude}
In this paper, we study the effect of team composition in the largest multiplayer online battle arena (MOBA) game: \kingofglory. 
We quantitatively show the varying effects of team composition on team performance (winning), team tenacity (surrender), and team rapport (abusive language use): 
although diverse teams tend to perform well and show tenacity in adversity, they are more likely to abuse when losing.
The double-edged influence of team diversity suggests the importance of 
balancing team composition.
We also examine how team composition influences individual behavior in abusive language use.
In addition to showing that a team is not the sum of independent individuals,
we contribute to the ``situation vs. personality'' debate and show that assassins abuse more because abusive players tend to choose assassins instead of players becoming abusive when choosing assassins.
Our work suggests that the gaming environment may be improved by adjusting team matching and preventing players from using team compositions that correlate with increased abusive language use. 

\para{Limitations.}
Although we control for roles in the analyses and further validate our results through prediction experiments, our study is observational and can be strengthened through experimental studies.
We would like to highlight that experimental studies on team compositions are non-trivial because of the great number of possible combinations.
Our work is also limited by the data that we have access to. 
Due to privacy issues, we do not have sensitive individual information like location and consumption record. 
Moreover, although \kingofglory is the largest MOBA game, the selection bias in our data may limit the generalizability of our findings.

\para{Future directions.}
Several promising directions arise from our work.
Our prediction experiments show that it is important to incorporate individual player information with role combinations.
It remains challenging to develop a holistic model that captures the interaction between individual members, learns novel representations for role combinations, and makes even more accurate predictions.
Furthermore, although MOBA games provide an ideal environment for understanding the effect of team composition, it is important to validate our findings in other scenarios, e.g., with implicit roles beyond gaming.

%% file: appendix.tex
\clearpage

\appendix
\section{Appendix}
\label{sec:appendix}
\subsection{Game information}
\tableref{tb:tag info} shows the detailed information contained in the game records.

\begin{table}[h]
	\centering
	\caption{The list of available information for each game.}
	\begin{tabular}{p{2.5cm} p{4.5cm}}
		\toprule
		Tag & Description \\
		\toprule
		\multicolumn{2}{c}{\textbf{Player's basic information}}\\
		$\#$grade-of-rank & Player's rank level, 0-26.\\
		$\#$heroid & Hero that player uses in this game.\\
		\hline
		\multicolumn{2}{c}{\textbf{Game's basic information}}\\
		$\#$game-time & UTC timestamp, game start time.\\
		$\#$duration & Time duration of the game.\\
		$\#$gameresult & Win/lose result for the player.\\
		$\#$surrender-tag & The tag of whether the game ends with surrendering.\\
		\hline
		\multicolumn{2}{c}{\textbf{In-game information}}\\
		$\#$kill-cnt & Number of kills.\\
		$\#$dead-cnt & Number of deaths.\\
		$\#$assist-cnt & Number of assistance kills.\\
		coin-cnt & Amount of coins.\\
		\hline
		\multicolumn{2}{c}{\textbf{Player's historical records}}\\
		rank-mvp & Number of mvp he/she got in his/her last 10 ranked games.\\
		$\#$ten-rank-win & The number of winning in his/her last 10 ranked games.\\
		$\#$ten-rank-total & Total number of previous 10 ranked games.\\
		\hline
		\multicolumn{2}{c}{\textbf{Chat information(if exists)}}\\
		$\#$timestamp & UTC timestamp, the time of one chat record.\\
		$\#$abuse-tag & The tag of whether this chat record includes abusing words.\\
		\toprule
	\end{tabular}
	\label{tb:tag info}
\end{table}

\subsection{Detailed Statistics}
We present team compositions with the highest (lowest) surrendering probability in \tableref{tbl:roles-surrender},
and those with the highest (lowest) abusing probability in \tableref{tbl:roles_abuse} .

\begin{table}[h]
	\centering
	\small
	\begin{tabular}{lrr}
		\toprule
		Team compositions  & Surrender prob. & Used frequency(\%)\\
		\midrule
		\multicolumn{3}{c}{team compositions with the highest surrendering rates} \\
		$\mage-\mage-\support-\support-\support$ & 84.6\% & $3.0 \times 10^{-4}$ \\
		$\mage-\mage-\mage-\support-\support$ & 82.4\% & $6.6 \times 10^{-4}$ \\
		$\assassin-\support-\support-\support-\warrior$ & 79.8\% & $9.7 \times 10^{-5}$\\
		\midrule
		\multicolumn{3}{c}{team compositions with the lowest surrendering rates} \\
		$\assassin-\assassin-\mage-\warrior-\warrior$ & 34.8\% & $4.6 \times 10^{-4}$ \\
		$\assassin-\assassin-\warrior-\warrior-\warrior$ & 34.3\% & $2.1 \times 10^{-5}$ \\
		$\assassin-\assassin-\marksman-\marksman-\warrior$ & 33.6\% & $1.5 \times 10^{-4}$\\
		\bottomrule
	\end{tabular}
	\caption{Team compositions with the highest and lowest surrender probability ($t \leq 11 min $).
	}
	\label{tbl:roles-surrender}
\end{table}

\begin{table}[h]
	\centering
	\small
	\begin{tabular}{lrr}
		\toprule
		Team compositions & Abusing prob. & Used frequency(\%)\\
		\midrule
		\multicolumn{3}{c}{team compositions with the highest abusing rates} \\
		$\assassin-\assassin-\assassin-\mage-\warrior$ & 56.2\%  & $2.2 \times 10^{-3}$ \\
		$\assassin-\assassin-\assassin-\marksman-\warrior$ & 55.9\% & $2.6 \times 10^{-3}$ \\
		$\assassin-\assassin-\assassin-\mage-\marksman$ & 55.6\%  & $7.2 \times 10^{-3}$\\
		\midrule
		\multicolumn{3}{c}{team compositions with the lowest abusing rates} \\
		$\mage-\mage-\mage-\support-\support$ & 30.6\% & $1.7 \times 10^{-3}$ \\
		$\support-\support-\support-\warrior-\warrior$ & 29.7\% & $1.5\times 10^{-5}$ \\
		$\mage-\support-\support-\support-\support$ & 28.7\% & $8.6 \times 10^{-5}$ \\
		\bottomrule
	\end{tabular}
	\caption{Team compositions with the highest and lowest abusing probability.}
	\label{tbl:roles_abuse}
\end{table}

%% file: arenamain.bbl

\begin{thebibliography}{49}


\ifx \showCODEN    \undefined \def \showCODEN     #1{\unskip}     \fi
\ifx \showDOI      \undefined \def \showDOI       #1{#1}\fi
\ifx \showISBNx    \undefined \def \showISBNx     #1{\unskip}     \fi
\ifx \showISBNxiii \undefined \def \showISBNxiii  #1{\unskip}     \fi
\ifx \showISSN     \undefined \def \showISSN      #1{\unskip}     \fi
\ifx \showLCCN     \undefined \def \showLCCN      #1{\unskip}     \fi
\ifx \shownote     \undefined \def \shownote      #1{#1}          \fi
\ifx \showarticletitle \undefined \def \showarticletitle #1{#1}   \fi
\ifx \showURL      \undefined \def \showURL       {\relax}        \fi
\providecommand\bibfield[2]{#2}
\providecommand\bibinfo[2]{#2}
\providecommand\natexlab[1]{#1}
\providecommand\showeprint[2][]{arXiv:#2}

\bibitem[\protect\citeauthoryear{Agarwala and Pearce}{Agarwala and
  Pearce}{2014}]%
        {agarwala2014learning}
\bibfield{author}{\bibinfo{person}{Atish Agarwala} {and}
  \bibinfo{person}{Michael Pearce}.} \bibinfo{year}{2014}\natexlab{}.
\newblock \bibinfo{booktitle}{\emph{Learning Dota 2 team compositions}}.
\newblock \bibinfo{type}{{T}echnical {R}eport}. \bibinfo{institution}{Technical
  report, Stanford University}.
\newblock


\bibitem[\protect\citeauthoryear{Anagnostopoulos, Becchetti, Castillo, Gionis,
  and Leonardi}{Anagnostopoulos et~al\mbox{.}}{2010}]%
        {Anagnostopoulos:ProceedingsOfCikm:2010}
\bibfield{author}{\bibinfo{person}{Aris Anagnostopoulos}, \bibinfo{person}{Luca
  Becchetti}, \bibinfo{person}{Carlos Castillo}, \bibinfo{person}{Aristides
  Gionis}, {and} \bibinfo{person}{Stefano Leonardi}.}
  \bibinfo{year}{2010}\natexlab{}.
\newblock \showarticletitle{{Power in unity: forming teams in large-scale
  community systems}}. In \bibinfo{booktitle}{\emph{CIKM}}.
\newblock


\bibitem[\protect\citeauthoryear{Anagnostopoulos, Becchetti, Castillo, Gionis,
  and Leonardi}{Anagnostopoulos et~al\mbox{.}}{2012}]%
        {anagnostopoulos2012online}
\bibfield{author}{\bibinfo{person}{Aris Anagnostopoulos}, \bibinfo{person}{Luca
  Becchetti}, \bibinfo{person}{Carlos Castillo}, \bibinfo{person}{Aristides
  Gionis}, {and} \bibinfo{person}{Stefano Leonardi}.}
  \bibinfo{year}{2012}\natexlab{}.
\newblock \showarticletitle{Online team formation in social networks}. In
  \bibinfo{booktitle}{\emph{WWW}}.
\newblock


\bibitem[\protect\citeauthoryear{Audas, Dobson, and Goddard}{Audas
  et~al\mbox{.}}{2002}]%
        {JournalOfEconomicsAndBusiness:2002}
\bibfield{author}{\bibinfo{person}{Rick Audas}, \bibinfo{person}{Stephen
  Dobson}, {and} \bibinfo{person}{John Goddard}.}
  \bibinfo{year}{2002}\natexlab{}.
\newblock \showarticletitle{{The impact of managerial change on team
  performance in professional sports}}.
\newblock \bibinfo{journal}{\emph{Journal of Economics and Business}}
  \bibinfo{volume}{54}, \bibinfo{number}{6} (\bibinfo{year}{2002}),
  \bibinfo{pages}{633--650}.
\newblock


\bibitem[\protect\citeauthoryear{Baum and Locke}{Baum and Locke}{2004}]%
        {baum2004relationship}
\bibfield{author}{\bibinfo{person}{J~Robert Baum} {and}
  \bibinfo{person}{Edwin~A Locke}.} \bibinfo{year}{2004}\natexlab{}.
\newblock \showarticletitle{The relationship of entrepreneurial traits, skill,
  and motivation to subsequent venture growth.}
\newblock \bibinfo{journal}{\emph{Journal of applied psychology}}
  \bibinfo{volume}{89}, \bibinfo{number}{4} (\bibinfo{year}{2004}),
  \bibinfo{pages}{587}.
\newblock


\bibitem[\protect\citeauthoryear{Belbin}{Belbin}{2011}]%
        {meredith2011management}
\bibfield{author}{\bibinfo{person}{R~Meredith Belbin}.}
  \bibinfo{year}{2011}\natexlab{}.
\newblock \showarticletitle{Management teams: Why they succeed or fail}.
\newblock \bibinfo{journal}{\emph{Human Resource Management International
  Digest}} \bibinfo{volume}{19}, \bibinfo{number}{3} (\bibinfo{year}{2011}).
\newblock


\bibitem[\protect\citeauthoryear{Belbin}{Belbin}{2012}]%
        {belbin2012team}
\bibfield{author}{\bibinfo{person}{R~Meredith Belbin}.}
  \bibinfo{year}{2012}\natexlab{}.
\newblock \bibinfo{booktitle}{\emph{Team roles at work}}.
\newblock \bibinfo{publisher}{Routledge}.
\newblock


\bibitem[\protect\citeauthoryear{Biddle}{Biddle}{1986}]%
        {biddle1986recent}
\bibfield{author}{\bibinfo{person}{Bruce~J Biddle}.}
  \bibinfo{year}{1986}\natexlab{}.
\newblock \showarticletitle{Recent developments in role theory}.
\newblock \bibinfo{journal}{\emph{Annual review of sociology}}
  \bibinfo{volume}{12}, \bibinfo{number}{1} (\bibinfo{year}{1986}),
  \bibinfo{pages}{67--92}.
\newblock


\bibitem[\protect\citeauthoryear{Bradner, Mark, and Hertel}{Bradner
  et~al\mbox{.}}{2005}]%
        {bradner2005team}
\bibfield{author}{\bibinfo{person}{Erin Bradner}, \bibinfo{person}{Gloria
  Mark}, {and} \bibinfo{person}{Tammie~D Hertel}.}
  \bibinfo{year}{2005}\natexlab{}.
\newblock \showarticletitle{Team size and technology fit: Participation,
  awareness, and rapport in distributed teams}.
\newblock \bibinfo{journal}{\emph{IEEE Transactions on Professional
  Communication}} \bibinfo{volume}{48}, \bibinfo{number}{1}
  (\bibinfo{year}{2005}), \bibinfo{pages}{68--77}.
\newblock


\bibitem[\protect\citeauthoryear{Brannick, Roach, and Salas}{Brannick
  et~al\mbox{.}}{1993}]%
        {micheal1993understanding}
\bibfield{author}{\bibinfo{person}{Michael~T. Brannick},
  \bibinfo{person}{Regina~M. Roach}, {and} \bibinfo{person}{Eduardo Salas}.}
  \bibinfo{year}{1993}\natexlab{}.
\newblock \showarticletitle{Understanding Team Performance: A Multimethod
  Study}.
\newblock \bibinfo{journal}{\emph{Human Performance}} \bibinfo{volume}{6},
  \bibinfo{number}{4} (\bibinfo{year}{1993}), \bibinfo{pages}{287--308}.
\newblock


\bibitem[\protect\citeauthoryear{Chen and Guestrin}{Chen and Guestrin}{2016}]%
        {chen2016xgboost}
\bibfield{author}{\bibinfo{person}{Tianqi Chen} {and} \bibinfo{person}{Carlos
  Guestrin}.} \bibinfo{year}{2016}\natexlab{}.
\newblock \showarticletitle{Xgboost: A scalable tree boosting system}. In
  \bibinfo{booktitle}{\emph{KDD}}.
\newblock


\bibitem[\protect\citeauthoryear{Cheng, Bernstein, Danescu-Niculescu-Mizil, and
  Leskovec}{Cheng et~al\mbox{.}}{2017}]%
        {cheng2017}
\bibfield{author}{\bibinfo{person}{Justin Cheng}, \bibinfo{person}{Michael
  Bernstein}, \bibinfo{person}{Cristian Danescu-Niculescu-Mizil}, {and}
  \bibinfo{person}{Jure Leskovec}.} \bibinfo{year}{2017}\natexlab{}.
\newblock \showarticletitle{Anyone can become a troll: Causes of trolling
  behavior in online discussions}. In \bibinfo{booktitle}{\emph{CSCW}}.
\newblock


\bibitem[\protect\citeauthoryear{Cheng, Danescu-Niculescu-Mizil, and
  Leskovec}{Cheng et~al\mbox{.}}{2015}]%
        {Cheng:ProceedingsOfIcwsm:2015}
\bibfield{author}{\bibinfo{person}{Justin Cheng}, \bibinfo{person}{Cristian
  Danescu-Niculescu-Mizil}, {and} \bibinfo{person}{Jure Leskovec}.}
  \bibinfo{year}{2015}\natexlab{}.
\newblock \showarticletitle{{Antisocial Behavior in Online Discussion
  Communities}}. In \bibinfo{booktitle}{\emph{ICWSM}}.
\newblock


\bibitem[\protect\citeauthoryear{Cohen and Bailey}{Cohen and Bailey}{1997}]%
        {susan1997what}
\bibfield{author}{\bibinfo{person}{Susan~G. Cohen} {and}
  \bibinfo{person}{Diane~E. Bailey}.} \bibinfo{year}{1997}\natexlab{}.
\newblock \showarticletitle{What Makes Teams Work: Group Effectiveness Research
  from the Shop Floor to the Executive Suite}.
\newblock \bibinfo{journal}{\emph{Journal of Management}} \bibinfo{volume}{23},
  \bibinfo{number}{3} (\bibinfo{year}{1997}), \bibinfo{pages}{239--290}.
\newblock


\bibitem[\protect\citeauthoryear{Conley and Perry}{Conley and Perry}{2013}]%
        {conley2013does}
\bibfield{author}{\bibinfo{person}{Kevin Conley} {and} \bibinfo{person}{Daniel
  Perry}.} \bibinfo{year}{2013}\natexlab{}.
\newblock \showarticletitle{How does he saw me? A recommendation engine for
  picking heroes in Dota 2}.
\newblock \bibinfo{journal}{\emph{Np, nd Web}}  \bibinfo{volume}{7}
  (\bibinfo{year}{2013}).
\newblock


\bibitem[\protect\citeauthoryear{Danescu-Niculescu-Mizil, Lee, Pang, and
  Kleinberg}{Danescu-Niculescu-Mizil et~al\mbox{.}}{2012}]%
        {Danescu-Niculescu-Mizil:2012:EPL:2187836.2187931}
\bibfield{author}{\bibinfo{person}{Cristian Danescu-Niculescu-Mizil},
  \bibinfo{person}{Lillian Lee}, \bibinfo{person}{Bo Pang}, {and}
  \bibinfo{person}{Jon Kleinberg}.} \bibinfo{year}{2012}\natexlab{}.
\newblock \showarticletitle{{Echoes of Power: Language Effects and Power
  Differences in Social Interaction}}. In \bibinfo{booktitle}{\emph{WWW}}.
\newblock


\bibitem[\protect\citeauthoryear{Dirks}{Dirks}{1999}]%
        {dirks1999effects}
\bibfield{author}{\bibinfo{person}{Kurt~T Dirks}.}
  \bibinfo{year}{1999}\natexlab{}.
\newblock \showarticletitle{The effects of interpersonal trust on work group
  performance.}
\newblock \bibinfo{journal}{\emph{Journal of applied psychology}}
  \bibinfo{volume}{84}, \bibinfo{number}{3} (\bibinfo{year}{1999}),
  \bibinfo{pages}{445}.
\newblock


\bibitem[\protect\citeauthoryear{Donellan, Lucas, and Fleeson}{Donellan
  et~al\mbox{.}}{2009}]%
        {person-situation-09}
\bibfield{editor}{\bibinfo{person}{M.~Brent Donellan},
  \bibinfo{person}{Richard~E. Lucas}, {and} \bibinfo{person}{William Fleeson}}
  (Eds.). \bibinfo{year}{2009}\natexlab{}.
\newblock \bibinfo{booktitle}{\emph{Personality and Assessment at Age 40:
  Reflections on the Past Person-Situation Debate and Emerging Directions of
  Future Person-Situation Integration [Special Issue]}}.
\newblock


\bibitem[\protect\citeauthoryear{Ducheneaut, Yee, Nickell, and
  Moore}{Ducheneaut et~al\mbox{.}}{2006}]%
        {ducheneaut2006alone}
\bibfield{author}{\bibinfo{person}{Nicolas Ducheneaut},
  \bibinfo{person}{Nicholas Yee}, \bibinfo{person}{Eric Nickell}, {and}
  \bibinfo{person}{Robert~J Moore}.} \bibinfo{year}{2006}\natexlab{}.
\newblock \showarticletitle{Alone together?: exploring the social dynamics of
  massively multiplayer online games}. In \bibinfo{booktitle}{\emph{Proceedings
  of the SIGCHI conference on Human Factors in computing systems}}. ACM,
  \bibinfo{pages}{407--416}.
\newblock


\bibitem[\protect\citeauthoryear{Ducheneaut, Yee, Nickell, and
  Moore}{Ducheneaut et~al\mbox{.}}{2007}]%
        {Ducheneaut:ProceedingsOfChi:2007}
\bibfield{author}{\bibinfo{person}{Nicolas Ducheneaut},
  \bibinfo{person}{Nicholas Yee}, \bibinfo{person}{Eric Nickell}, {and}
  \bibinfo{person}{Robert~J. Moore}.} \bibinfo{year}{2007}\natexlab{}.
\newblock \showarticletitle{{The Life and Death of Online Gaming Communities}}.
  In \bibinfo{booktitle}{\emph{CHI}}.
\newblock


\bibitem[\protect\citeauthoryear{Duckworth, Peterson, Matthews, and
  Kelly}{Duckworth et~al\mbox{.}}{2007}]%
        {duckworth2007grit}
\bibfield{author}{\bibinfo{person}{Angela~L Duckworth},
  \bibinfo{person}{Christopher Peterson}, \bibinfo{person}{Michael~D Matthews},
  {and} \bibinfo{person}{Dennis~R Kelly}.} \bibinfo{year}{2007}\natexlab{}.
\newblock \showarticletitle{Grit: perseverance and passion for long-term
  goals.}
\newblock \bibinfo{journal}{\emph{Journal of personality and social
  psychology}} \bibinfo{volume}{92}, \bibinfo{number}{6}
  (\bibinfo{year}{2007}), \bibinfo{pages}{1087}.
\newblock


\bibitem[\protect\citeauthoryear{Freeman}{Freeman}{2018}]%
        {freeman2018multiplayer}
\bibfield{author}{\bibinfo{person}{Guo Freeman}.}
  \bibinfo{year}{2018}\natexlab{}.
\newblock \bibinfo{booktitle}{\emph{Multiplayer Online Games: Origins, Players,
  and Social Dynamics}}.
\newblock \bibinfo{publisher}{AK Peters/CRC Press}.
\newblock


\bibitem[\protect\citeauthoryear{Honeycutt}{Honeycutt}{2005}]%
        {honeycutt2005hazing}
\bibfield{author}{\bibinfo{person}{Courtenay Honeycutt}.}
  \bibinfo{year}{2005}\natexlab{}.
\newblock \showarticletitle{Hazing as a process of boundary maintenance in an
  online community}.
\newblock \bibinfo{journal}{\emph{Journal of computer-mediated communication}}
  \bibinfo{volume}{10}, \bibinfo{number}{2} (\bibinfo{year}{2005}),
  \bibinfo{pages}{JCMC1021}.
\newblock


\bibitem[\protect\citeauthoryear{Hudson and Cairns}{Hudson and Cairns}{2014}]%
        {hudson2014measuring}
\bibfield{author}{\bibinfo{person}{Matthew Hudson} {and} \bibinfo{person}{Paul
  Cairns}.} \bibinfo{year}{2014}\natexlab{}.
\newblock \showarticletitle{Measuring social presence in team-based digital
  games}.
\newblock \bibinfo{journal}{\emph{Interacting with Presence: HCI and the Sense
  of Presence in Computer-mediated Environments}} (\bibinfo{year}{2014}),
  \bibinfo{pages}{83}.
\newblock


\bibitem[\protect\citeauthoryear{Kenrick and Funder}{Kenrick and
  Funder}{1988}]%
        {Kenrick:AmericanPsychologist:1988}
\bibfield{author}{\bibinfo{person}{Douglas~T Kenrick} {and}
  \bibinfo{person}{David~C Funder}.} \bibinfo{year}{1988}\natexlab{}.
\newblock \showarticletitle{{Profiting from controversy: Lessons from the
  person-situation debate}}.
\newblock \bibinfo{journal}{\emph{American Psychologist}} \bibinfo{volume}{43},
  \bibinfo{number}{1} (\bibinfo{year}{1988}), \bibinfo{pages}{23}.
\newblock


\bibitem[\protect\citeauthoryear{Khatib, Cooper, Tyka, Xu, Makedon,
  Popovi{\'c}, and Baker}{Khatib et~al\mbox{.}}{2011}]%
        {khatib2011algorithm}
\bibfield{author}{\bibinfo{person}{Firas Khatib}, \bibinfo{person}{Seth
  Cooper}, \bibinfo{person}{Michael~D Tyka}, \bibinfo{person}{Kefan Xu},
  \bibinfo{person}{Ilya Makedon}, \bibinfo{person}{Zoran Popovi{\'c}}, {and}
  \bibinfo{person}{David Baker}.} \bibinfo{year}{2011}\natexlab{}.
\newblock \showarticletitle{Algorithm discovery by protein folding game
  players}.
\newblock \bibinfo{journal}{\emph{the National Academy of Sciences}}
  \bibinfo{volume}{108}, \bibinfo{number}{47} (\bibinfo{year}{2011}),
  \bibinfo{pages}{18949--18953}.
\newblock


\bibitem[\protect\citeauthoryear{Kim, Engel, Woolley, Lin, McArthur, and
  Malone}{Kim et~al\mbox{.}}{2017}]%
        {Kim:CommunicationsOfTheAcm:}
\bibfield{author}{\bibinfo{person}{Young~Ji Kim}, \bibinfo{person}{David
  Engel}, \bibinfo{person}{Anita~Williams Woolley}, \bibinfo{person}{Jeffrey
  Yu-Ting Lin}, \bibinfo{person}{Naomi McArthur}, {and}
  \bibinfo{person}{Thomas~W. Malone}.} \bibinfo{year}{2017}\natexlab{}.
\newblock \showarticletitle{{What Makes a Strong Team?: Using Collective
  Intelligence to Predict Team Performance in League of Legends}}. In
  \bibinfo{booktitle}{\emph{CSCW}}.
\newblock


\bibitem[\protect\citeauthoryear{Kuss, Louws, and Wiers}{Kuss
  et~al\mbox{.}}{2012}]%
        {kuss2012online}
\bibfield{author}{\bibinfo{person}{Daria~J Kuss}, \bibinfo{person}{Jorik
  Louws}, {and} \bibinfo{person}{Reinout~W Wiers}.}
  \bibinfo{year}{2012}\natexlab{}.
\newblock \showarticletitle{Online gaming addiction? Motives predict addictive
  play behavior in massively multiplayer online role-playing games}.
\newblock \bibinfo{journal}{\emph{Cyberpsychology, Behavior, and Social
  Networking}} \bibinfo{volume}{15}, \bibinfo{number}{9}
  (\bibinfo{year}{2012}), \bibinfo{pages}{480--485}.
\newblock


\bibitem[\protect\citeauthoryear{Kwak, Blackburn, and Han}{Kwak
  et~al\mbox{.}}{2015}]%
        {Kwak2015}
\bibfield{author}{\bibinfo{person}{Haewoon Kwak}, \bibinfo{person}{Jeremy
  Blackburn}, {and} \bibinfo{person}{Seungyeop Han}.}
  \bibinfo{year}{2015}\natexlab{}.
\newblock \showarticletitle{Exploring Cyberbullying and Other Toxic Behavior in
  Team Competition Online Games}. In \bibinfo{booktitle}{\emph{CHI}}.
\newblock


\bibitem[\protect\citeauthoryear{Lappas, Liu, and Terzi}{Lappas
  et~al\mbox{.}}{2009}]%
        {Lappas:ProceedingsOfKdd:2009}
\bibfield{author}{\bibinfo{person}{Theodoros Lappas}, \bibinfo{person}{Kun
  Liu}, {and} \bibinfo{person}{Evimaria Terzi}.}
  \bibinfo{year}{2009}\natexlab{}.
\newblock \showarticletitle{{Finding a team of experts in social networks}}. In
  \bibinfo{booktitle}{\emph{KDD}}.
\newblock


\bibitem[\protect\citeauthoryear{Li, Tong, Cao, Ehrlich, Lin, and Buchler}{Li
  et~al\mbox{.}}{2017}]%
        {Li:IeeeTransKnowlDataEng:2017}
\bibfield{author}{\bibinfo{person}{Liangyue Li}, \bibinfo{person}{Hanghang
  Tong}, \bibinfo{person}{Nan Cao}, \bibinfo{person}{Kate Ehrlich},
  \bibinfo{person}{Yu-Ru Lin}, {and} \bibinfo{person}{Norbou Buchler}.}
  \bibinfo{year}{2017}\natexlab{}.
\newblock \showarticletitle{{Enhancing Team Composition in Professional
  Networks: Problem Definitions and Fast Solutions}}.
\newblock \bibinfo{journal}{\emph{IEEE Trans Knowl Data Eng}}
  \bibinfo{volume}{29}, \bibinfo{number}{3} (\bibinfo{year}{2017}),
  \bibinfo{pages}{613--626}.
\newblock


\bibitem[\protect\citeauthoryear{Losup, Van De~Bovenkamp, Shen, Jia, and
  Kuipers}{Losup et~al\mbox{.}}{2014}]%
        {losup2014analyzing}
\bibfield{author}{\bibinfo{person}{Alexandru Losup}, \bibinfo{person}{Ruud Van
  De~Bovenkamp}, \bibinfo{person}{Siqi Shen}, \bibinfo{person}{Adele~Lu Jia},
  {and} \bibinfo{person}{Fernando Kuipers}.} \bibinfo{year}{2014}\natexlab{}.
\newblock \showarticletitle{Analyzing implicit social networks in multiplayer
  online games}.
\newblock \bibinfo{journal}{\emph{IEEE Internet Computing}} \bibinfo{number}{3}
  (\bibinfo{year}{2014}), \bibinfo{pages}{36--44}.
\newblock


\bibitem[\protect\citeauthoryear{Norms and Niebuhr}{Norms and Niebuhr}{1980}]%
        {norms1980}
\bibfield{author}{\bibinfo{person}{Dwight~R Norms} {and}
  \bibinfo{person}{Robert~E Niebuhr}.} \bibinfo{year}{1980}\natexlab{}.
\newblock \showarticletitle{Group variables and gaming success}.
\newblock \bibinfo{journal}{\emph{Simulation \& Games}} \bibinfo{volume}{11},
  \bibinfo{number}{3} (\bibinfo{year}{1980}), \bibinfo{pages}{301--312}.
\newblock


\bibitem[\protect\citeauthoryear{Parker}{Parker}{1990}]%
        {parker1990team}
\bibfield{author}{\bibinfo{person}{Glenn~M Parker}.}
  \bibinfo{year}{1990}\natexlab{}.
\newblock \bibinfo{booktitle}{\emph{Team players and teamwork}}.
\newblock \bibinfo{publisher}{Jossey-Bass San Francisco, CA}.
\newblock


\bibitem[\protect\citeauthoryear{Partington and Harris}{Partington and
  Harris}{1999}]%
        {partington1999team}
\bibfield{author}{\bibinfo{person}{David Partington} {and}
  \bibinfo{person}{Hilary Harris}.} \bibinfo{year}{1999}\natexlab{}.
\newblock \showarticletitle{Team role balance and team performance: an
  empirical study}.
\newblock \bibinfo{journal}{\emph{Journal of Management Development}}
  \bibinfo{volume}{18}, \bibinfo{number}{8} (\bibinfo{year}{1999}),
  \bibinfo{pages}{694--705}.
\newblock


\bibitem[\protect\citeauthoryear{Petter}{Petter}{2017}]%
        {petter2017}
\bibfield{author}{\bibinfo{person}{Stacie Petter}.}
  \bibinfo{year}{2017}\natexlab{}.
\newblock \showarticletitle{More Than Child's Play: Embracing the Study of
  Online Gaming in Information Systems Research}.
\newblock \bibinfo{journal}{\emph{SIGMIS Database}} \bibinfo{volume}{48},
  \bibinfo{number}{4} (\bibinfo{date}{Nov.} \bibinfo{year}{2017}),
  \bibinfo{pages}{9--13}.
\newblock
\showISSN{0095-0033}


\bibitem[\protect\citeauthoryear{Pobiedina, Neidhardt, Calatrava~Moreno, and
  Werthner}{Pobiedina et~al\mbox{.}}{2013a}]%
        {pobiedina2013ranking}
\bibfield{author}{\bibinfo{person}{Nataliia Pobiedina}, \bibinfo{person}{Julia
  Neidhardt}, \bibinfo{person}{Maria del~Carmen Calatrava~Moreno}, {and}
  \bibinfo{person}{Hannes Werthner}.} \bibinfo{year}{2013}\natexlab{a}.
\newblock \showarticletitle{Ranking factors of team success}. In
  \bibinfo{booktitle}{\emph{Proceedings of the 22nd International Conference on
  World Wide Web}}. ACM, \bibinfo{pages}{1185--1194}.
\newblock


\bibitem[\protect\citeauthoryear{Pobiedina, Neidhardt, Moreno, Grad-Gyenge, and
  Werthner}{Pobiedina et~al\mbox{.}}{2013b}]%
        {pobiedina2013successful}
\bibfield{author}{\bibinfo{person}{Nataliia Pobiedina}, \bibinfo{person}{Julia
  Neidhardt}, \bibinfo{person}{Maria del Carmen~Calatrava Moreno},
  \bibinfo{person}{Laszlo Grad-Gyenge}, {and} \bibinfo{person}{Hannes
  Werthner}.} \bibinfo{year}{2013}\natexlab{b}.
\newblock \showarticletitle{On successful team formation: Statistical analysis
  of a multiplayer online game}. In \bibinfo{booktitle}{\emph{Business
  Informatics (CBI), 2013 IEEE 15th Conference on}}. IEEE,
  \bibinfo{pages}{55--62}.
\newblock


\bibitem[\protect\citeauthoryear{Salas, Cooke, and Rosen}{Salas
  et~al\mbox{.}}{2008}]%
        {salas2008teams}
\bibfield{author}{\bibinfo{person}{Eduardo Salas}, \bibinfo{person}{Nancy~J
  Cooke}, {and} \bibinfo{person}{Michael~A Rosen}.}
  \bibinfo{year}{2008}\natexlab{}.
\newblock \showarticletitle{On teams, teamwork, and team performance:
  Discoveries and developments}.
\newblock \bibinfo{journal}{\emph{Human factors}} \bibinfo{volume}{50},
  \bibinfo{number}{3} (\bibinfo{year}{2008}), \bibinfo{pages}{540--547}.
\newblock


\bibitem[\protect\citeauthoryear{Sapienza, Goyal, and Ferrara}{Sapienza
  et~al\mbox{.}}{2018}]%
        {sapienza2018deep}
\bibfield{author}{\bibinfo{person}{Anna Sapienza}, \bibinfo{person}{Palash
  Goyal}, {and} \bibinfo{person}{Emilio Ferrara}.}
  \bibinfo{year}{2018}\natexlab{}.
\newblock \showarticletitle{Deep Neural Networks for Optimal Team Composition}.
\newblock \bibinfo{journal}{\emph{arXiv preprint arXiv:1805.03285}}
  (\bibinfo{year}{2018}).
\newblock


\bibitem[\protect\citeauthoryear{Senior}{Senior}{1997}]%
        {barbara1997}
\bibfield{author}{\bibinfo{person}{Barbara Senior}.}
  \bibinfo{year}{1997}\natexlab{}.
\newblock \showarticletitle{Team roles and team performance: Is there 'really'
  a link?}
\newblock \bibinfo{journal}{\emph{Journal of Occupational and Organizational
  Psychology}} \bibinfo{volume}{70}, \bibinfo{number}{3}
  (\bibinfo{year}{1997}), \bibinfo{pages}{241--258}.
\newblock
\showISSN{2044-8325}


\bibitem[\protect\citeauthoryear{Stetina, Kothgassner, Lehenbauer, and
  Kryspin-Exner}{Stetina et~al\mbox{.}}{2011}]%
        {birgit2011}
\bibfield{author}{\bibinfo{person}{Birgit~U. Stetina},
  \bibinfo{person}{Oswald~D. Kothgassner}, \bibinfo{person}{Mario Lehenbauer},
  {and} \bibinfo{person}{Ilse Kryspin-Exner}.} \bibinfo{year}{2011}\natexlab{}.
\newblock \showarticletitle{Beyond the fascination of online-games: Probing
  addictive behavior and depression in the world of online-gaming}.
\newblock \bibinfo{journal}{\emph{Computers in Human Behavior}}
  \bibinfo{volume}{27}, \bibinfo{number}{1} (\bibinfo{year}{2011}),
  \bibinfo{pages}{473 -- 479}.
\newblock
\showISSN{0747-5632}
\newblock
\shownote{Current Research Topics in Cognitive Load Theory.}


\bibitem[\protect\citeauthoryear{Stewart, Fulmer, and Barrick}{Stewart
  et~al\mbox{.}}{2005}]%
        {stewart2005}
\bibfield{author}{\bibinfo{person}{Greg~L. Stewart}, \bibinfo{person}{Ingrid~S.
  Fulmer}, {and} \bibinfo{person}{Murray~R. Barrick}.}
  \bibinfo{year}{2005}\natexlab{}.
\newblock \showarticletitle{An exploration of member roles as a multilevel
  linking mechanism for individual traits and team outcomes}.
\newblock \bibinfo{journal}{\emph{Personnel Psychology}} \bibinfo{volume}{58},
  \bibinfo{number}{2} (\bibinfo{year}{2005}), \bibinfo{pages}{343--365}.
\newblock
\showISSN{1744-6570}


\bibitem[\protect\citeauthoryear{Szell and Thurner}{Szell and Thurner}{2010}]%
        {szell2010measuring}
\bibfield{author}{\bibinfo{person}{Michael Szell} {and} \bibinfo{person}{Stefan
  Thurner}.} \bibinfo{year}{2010}\natexlab{}.
\newblock \showarticletitle{Measuring social dynamics in a massive multiplayer
  online game}.
\newblock \bibinfo{journal}{\emph{Social networks}} \bibinfo{volume}{32},
  \bibinfo{number}{4} (\bibinfo{year}{2010}), \bibinfo{pages}{313--329}.
\newblock


\bibitem[\protect\citeauthoryear{Wuchty, Jones, and Uzzi}{Wuchty
  et~al\mbox{.}}{2007}]%
        {Wuchty:Science:2007}
\bibfield{author}{\bibinfo{person}{Stefan Wuchty}, \bibinfo{person}{Benjamin~F
  Jones}, {and} \bibinfo{person}{Brian Uzzi}.} \bibinfo{year}{2007}\natexlab{}.
\newblock \showarticletitle{{The Increasing Dominance of Teams in Production of
  Knowledge}}.
\newblock \bibinfo{journal}{\emph{Science}} \bibinfo{volume}{316},
  \bibinfo{number}{5827} (\bibinfo{year}{2007}), \bibinfo{pages}{1036--1039}.
\newblock


\bibitem[\protect\citeauthoryear{Xia, Wang, and Zhou}{Xia
  et~al\mbox{.}}{2017}]%
        {xia2017contributes}
\bibfield{author}{\bibinfo{person}{Bang Xia}, \bibinfo{person}{Huiwen Wang},
  {and} \bibinfo{person}{Ronggang Zhou}.} \bibinfo{year}{2017}\natexlab{}.
\newblock \showarticletitle{What Contributes to Success in MOBA Games? An
  Empirical Study of Defense of the Ancients 2}.
\newblock \bibinfo{journal}{\emph{Games and Culture}} (\bibinfo{year}{2017}),
  \bibinfo{pages}{1555412017710599}.
\newblock


\bibitem[\protect\citeauthoryear{Yang, Halfaker, Kraut, and Hovy}{Yang
  et~al\mbox{.}}{2016}]%
        {Yang:ProceedingsOfIcwsm:2016}
\bibfield{author}{\bibinfo{person}{Diyi Yang}, \bibinfo{person}{Aaron
  Halfaker}, \bibinfo{person}{Robert Kraut}, {and} \bibinfo{person}{Eduard
  Hovy}.} \bibinfo{year}{2016}\natexlab{}.
\newblock \showarticletitle{{Who Did What: Editor Role Identification in
  Wikipedia}}. In \bibinfo{booktitle}{\emph{ICWSM}}.
\newblock


\bibitem[\protect\citeauthoryear{Yang, Harrison, and Roberts}{Yang
  et~al\mbox{.}}{2014}]%
        {yang2014}
\bibfield{author}{\bibinfo{person}{Pu Yang}, \bibinfo{person}{Brent~E
  Harrison}, {and} \bibinfo{person}{David~L Roberts}.}
  \bibinfo{year}{2014}\natexlab{}.
\newblock \showarticletitle{Identifying patterns in combat that are predictive
  of success in MOBA games.}. In \bibinfo{booktitle}{\emph{FDG}}.
\newblock


\bibitem[\protect\citeauthoryear{Yang, Tang, Leung, Sun, Chen, Li, and
  Yang}{Yang et~al\mbox{.}}{2015}]%
        {yang2015rain}
\bibfield{author}{\bibinfo{person}{Yang Yang}, \bibinfo{person}{Jie Tang},
  \bibinfo{person}{Cane Wing-ki Leung}, \bibinfo{person}{Yizhou Sun},
  \bibinfo{person}{Qicong Chen}, \bibinfo{person}{Juanzi Li}, {and}
  \bibinfo{person}{Qiang Yang}.} \bibinfo{year}{2015}\natexlab{}.
\newblock \showarticletitle{RAIN: Social Role-Aware Information Diffusion}. In
  \bibinfo{booktitle}{\emph{AAAI'15}}. \bibinfo{pages}{367--373}.
\newblock


\end{thebibliography}
